\documentclass[fleqn,usenatbib]{mnras}

\usepackage{newtxtext,newtxmath}

\usepackage[T1]{fontenc}

\DeclareRobustCommand{\VAN}[3]{#2}
\let\VANthebibliography\thebibliography
\def\thebibliography{\DeclareRobustCommand{\VAN}[3]{##3}\VANthebibliography}

\usepackage{graphicx}	
\usepackage{amsmath}
\usepackage{ulem}

\usepackage{orcidlink}
\usepackage{multirow}
\usepackage[capitalize]{cleveref}
\usepackage{xspace}
\usepackage{subcaption} 

\newcommand{\borg}{\texttt{BORG}\xspace}

\newcommand{\tmpp}{\texttt{2M++}\xspace}
\newcommand{\flamingo}{\texttt{FLAMINGO}\xspace}
\newcommand{\lcdm}{$\Lambda$CDM\xspace}
\newcommand{\msol}{M$_\odot$\xspace}
\newcommand{\msolh}{$h^{-1}$~M$_\odot$\xspace}
\newcommand{\mpch}{$h^{-1}$~Mpc\xspace}

\newcommand{\dquotes}[1]{``#1''}
\newcommand{\squotes}[1]{`#1'}
\newcommand{\masscrit}{$M_{\mathrm{200,crit}}$\xspace}
\newcommand{\rcrit}{$r_{\mathrm{200,crit}}$\xspace}
\newcommand{\tentoninety}{$10^{\mathrm{th}}-90^{\mathrm{th}}$\xspace}
\newcommand{\manticore}{\texttt{Manticore}\xspace}
\newcommand{\manticorelocal}{\texttt{Manticore-Local}\xspace}
\newcommand{\manticorelocallo}{\texttt{Manticore-Local-Lo}\xspace}
\newcommand{\manticoredeep}{\texttt{Manticore-Deep}\xspace}
\newcommand{\manticoremini}{\texttt{Manticore-Mini}\xspace}

\Crefname{figure}{Figure}{Figures}
\Crefname{equation}{Equation}{Equations}
\Crefname{table}{Table}{Tables}
\Crefname{section}{Section}{Sections}

\title[\manticorelocal]{The \manticore Project I: a digital twin of our cosmic neighbourhood from Bayesian field-level analysis}

\author[S. McAlpine et al.]{
Stuart McAlpine\,\textsuperscript{\orcidlink{0000-0002-8286-7809}},$^{\,1}$\thanks{E-mail: \url{stuart.mcalpine@fysik.su.se}}
Jens Jasche,$^{\,1}$
Metin Ata,$^{\,1,2}$
Guilhem Lavaux,$^{\,3}$
Richard Stiskalek,$^{\,4,5}$
\newauthor
\hspace{2pt}Carlos S. Frenk,$^{\,6}$
and Adrian Jenkins$^{\,6}$
\\
$^{1}$The Oskar Klein Centre, Department of Physics, Stockholm University, Albanova University Center, 106 91 Stockholm, Sweden\\
$^{2}$Center for Gravitational Physics and Quantum Information, Yukawa Institute for Theoretical Physics, Kyoto University, Kyoto 606-8502, Japan\\
$^{3}$CNRS \& Sorbonne Université, Institut d'Astrophysique de Paris (IAP), UMR 7095, 98 bis bd Arago, F-75014 Paris, France\\
$^{4}$Department of Physics, University of Oxford, Denys Wilkinson Building, Keble Road, Oxford, OX1 3RH, UK\\
$^{5}$Center for Computational Astrophysics, Flatiron Institute, 162 5th Ave, New York, NY 10010, USA\\
$^{6}$Institute for Computational Cosmology, Durham University, South Road, Durham DH1 3LE, United Kingdom
}

\date{Accepted XXX. Received YYY; in original form ZZZ}

\pubyear{2015}

\begin{document}
\label{firstpage}
\pagerange{\pageref{firstpage}--\pageref{lastpage}}
\maketitle

\begin{abstract}
We present the first results from the \manticore project, dubbed \manticorelocal, a suite of Bayesian constrained simulations of the nearby Universe, generated by fitting a physical structure formation model to the \tmpp\ galaxy catalogue using the \borg\ algorithm. This field-level inference yields physically consistent realizations of cosmic structure, leveraging a nonlinear gravitational solver, a refined galaxy bias model, and physics-informed priors. The \manticorelocal\ posterior realizations evolve within a parent cosmological volume statistically consistent with \lcdm, demonstrated through extensive posterior predictive tests of power spectra, bispectra, initial condition Gaussianity, and the halo mass function. The inferred local supervolume ($R<200$~Mpc, or $z \lesssim 0.05$) shows no significant deviation from cosmological expectations; notably, we find no evidence for a large local underdensity, with the mean density suppressed by only $\approx 5$\% relative to the cosmic mean. Our model identifies high-significance counterparts for fourteen prominent galaxy clusters—including Virgo, Coma, and Perseus—each within one degree of its observed sky position. Across the posterior ensemble, these counterparts are consistently detected with 2--4$\sigma$ significance, and their reconstructed masses and redshifts agree closely with observational estimates, confirming the inference's spatial and dynamical fidelity. The peculiar velocity field recovered by \manticorelocal\ achieves the highest Bayesian evidence across five independent datasets, surpassing state-of-the-art nonlinear models, linear theory, Wiener filtering, and machine learning approaches. Unlike methods yielding only point estimates or using simplified dynamics, \manticorelocal\ provides a full Bayesian posterior over cosmic structure and evolution, enabling rigorous uncertainty quantification. These results establish \manticorelocal\ as the most advanced constrained realization suite of the Local Universe to date, offering a robust statistical foundation for future studies of galaxy formation, velocity flows, and environmental dependencies in our cosmic neighbourhood.

\end{abstract}




\section{Introduction}
\label{sect:introduction}

The nearby Universe, characterized by prominent galaxy clusters, interconnected filaments, and expansive cosmic voids, serves as a unique laboratory for exploring the physical processes governing the formation and evolution of cosmic structures \citep{2006Sci...313..311B}. With observational data continually improving and cosmological tensions becoming increasingly pronounced, developing accurate and predictive models of our local cosmic environment is essential \citep{2022NewAR..9501659P,2022JHEAp..34...49A,2022AnPhy.44769159P,2023CQGra..40i4001A}.

Among these tensions, the discrepancy in the Hubble constant ($H_0$) between local distance-ladder measurements and cosmic microwave background (CMB) inferences remains particularly significant \citep{Planck2018,2021CQGra..38o3001D,Riess2022}. However, other anomalies complicate the cosmological picture: claims of a substantial local underdensity extending to $\sim$150 Mpc may bias local measurements of cosmic expansion \citep{Keenan2013, Whitbourn2014}, and unexpectedly coherent galaxy bulk flows have been observed on scales that challenge \lcdm predictions \citep{Watkins2009,2015MNRAS.449.4494H,2023MNRAS.524.1885W}. Additionally, hemispherical asymmetries in large-scale structure raise new questions about cosmic isotropy and the statistical homogeneity of the Universe \citep{2012JCAP...10..036M,Bengaly2018,Peebles2022}. Taken together, these discrepancies suggest either subtle observational systematics or the need for physics beyond \lcdm.

To evaluate such possibilities, we require cosmological models that satisfy a dual criterion: they must accurately reconstruct the specific matter distribution surrounding us, and, simultaneously reproduce the complex, nonlinear dynamics governing structure formation. Achieving both is essential for distinguishing local cosmic peculiarities from signatures of new physics and for connecting detailed observations to the broader cosmological context.

Historically, analyses of large-scale structure have relied on summary statistics—especially the two-point correlation function and power spectrum—to characterize clustering in an ensemble-averaged sense \citep{2004ApJ...606..702T,2022PhRvD.106j3530P}. While effective for constraining parameters in the nearly Gaussian early Universe, these tools are blind to the rich non-Gaussian morphology of the evolved cosmic web—its filaments, walls, voids, and anisotropies—which encode critical information at the field level \citep{2024arXiv240502252B,2024PhRvL.133v1006N,2019MNRAS.486.5061R}.

Bayesian field-level inference addresses this limitation by reconstructing the full three-dimensional matter density field from galaxy observations, treating initial conditions as parameters constrained by a forward model of structure formation. This approach naturally retains phase information and higher-order statistics, enabling reconstructions that are physically consistent and uncertainty-aware across linear and nonlinear regimes \citep{Jasche2013}. As part of the broader family of constrained simulations, field-level inference offers the ability to recover initial density fields and evolve them via $N$-body simulations, producing a physically coherent picture of the local Universe.

Multiple frameworks have emerged for constructing constrained simulations of the Local Universe. Beyond the Aquila Consortium’s \borg\ algorithm \citep{Jasche2013, Lavaux2016, Jasche2019, Stopyra2024_COLA}, which performs full Bayesian field-level inference from galaxy surveys, key efforts include the $\textit{Constrained Local UniversE Simulations}$ (CLUES) project \citep{Gottloeber2010,Sorce2016a}, based on the Hoffman-Ribak method \citep{Hoffman1991, Klypin2003} and Cosmicflows velocity constraints; the SLOW and $\textit{Exploring the Local Universe with reConstructed Initial Density field}$ (ELUCID) simulations \citep{Wang2016, 2022ApJ...936...11L,Dolag2023}, full-physics constrained hydrodynamical simulations; the Local Universe Model (LUM) \citep{Pfeifer2023}, which provides a statistical framework to evaluate constrained simulations against observed cluster distributions; and the $\textit{Constrained Simulations in BORG}$ (CSiBORG) suite \citep{2021PhRvD.103b3523B}, which resimulates \borg-inferred initial conditions at high resolution. 

These projects have been applied to a wide range of science goals, including modelling the local cosmic flow field and its peculiar velocities \citep{2024A&A...687A..85S,Stiskalek2025, 2025arXiv250109573L}, constraining modified gravity and fifth-force scenarios \citep{2018PhRvD..98h3010D,2021PhRvD.103b3523B}, probing void statistics and halo–void duality \citep{2022MNRAS.511L..45D}, examining galaxy alignments and environmental quenching \citep{2023arXiv230103581T,2025arXiv250314732G}, directly constraining galaxy formation physics \citep{Sorce2021,2024ApJ...966..236L}, recovering the dynamical formation history of the local group \citep{Carlesi2016,2024A&A...691A.348W} and validating the consistency and signatures of local structure with \lcdm expectations \citep{Libeskind2020,SLOW2024,2024A&A...692A.180J,2024arXiv241208708S}. Together, these efforts underscore the growing importance of constrained realizations as a tool for precision cosmology in the nearby Universe.

The Aquila Consortium has been at the forefront of developing field-level inference methods and their application to data. Using the \borg algorithm, Aquila has produced high-fidelity reconstructions of the local density and velocity fields \citep{2018MNRAS.474.3152D,2022MNRAS.511L..45D,2022MNRAS.512.5823M,2024MNRAS.534.3120S}, and demonstrated their value in cosmological parameter inference, bias modelling, and the resolution of degeneracies \citep{2019MNRAS.483L..64D,2019A&A...630A.151P,2020PhRvD.102j4060D,2021PhRvD.104j3516B,2022PhRvD.106j3526B,2024MNRAS.535.1258D}.

In this work, we introduce \manticore, a new Bayesian field-level inference model building upon and extending the \borg framework. \manticore incorporates a more flexible galaxy bias prescription, a generalized Poisson likelihood, and physics-informed priors that enforce—but do not rigidly impose—consistency with \lcdm. When applied to the \tmpp galaxy catalogue, this model yields a suite of constrained posterior simulations, which we term \manticorelocal, representing a high-resolution, physically self-consistent reconstruction of the Local Universe.

The \manticorelocal posterior ensemble offers significant improvements over prior constrained simulations. It accurately recovers the spatial positions, masses, and velocities of major galaxy clusters, yields a peculiar velocity field with the highest Bayesian evidence across five independent datasets, and reproduces the expected power spectrum, bispectrum, and halo mass function of a $\Lambda$CDM cosmology. These results position \manticorelocal\ as the most advanced constrained realization suite of our cosmic neighbourhood to date, enabling rigorous tests of cosmological models in the fully nonlinear regime.

The remainder of this paper is structured as follows. Section~\ref{sect:method} describes the observational data, the inference configuration used to produce the \manticorelocal\ ensemble, and the procedure for posterior resimulation. Section~\ref{sect:results} presents our main findings, including consistency tests with $\Lambda$CDM, analyses of the local density and velocity fields, and comparisons with the observed cluster population. We discuss our results in the context of prior work in Section~\ref{sect:discussion}, and conclude in Section~\ref{sect:conclusions}.

\section{Method}
\label{sect:method}

To construct a statistically faithful model of the local Universe, we adopt a two-stage methodology. First, we apply the \borg algorithm to infer the posterior distribution of the initial Gaussian white noise field—the latent structure underlying present-day observables. Throughout this work, we refer to this field as either the white noise field or the initial conditions. Second, we generate an ensemble of posterior realizations (henceforth, resimulations) by evolving independent samples from this posterior forward in time using an accurate $N$-body solver. This yields a suite of constrained simulations that match both the statistical properties and the dynamical history of the observed matter distribution.

This section details the constraints from the 2M++ catalogue that serve as input for the inference (\cref{sect:2mpp}), the setup of the Bayesian inference framework, the \borg algorithm, used in this study (\cref{sect:inference_setup}), and the procedure for generating the posterior resimulations (\cref{sect:posterior_resimulations}).

\subsection{The \tmpp galaxy catalogue}
\label{sect:2mpp}

As data constraints, we utilize galaxy count data from the \tmpp catalogue, which comprises photometric targets from the 2-Micron All-Sky Survey Extended Source Catalogue \citep[2MASS-XSC;][]{Huchra2012}, supplemented with spectroscopic redshifts from the 2MASS Redshift Survey \citep[2MRS;][]{Huchra2012}, the 6-Degree Field Galaxy Redshift Survey \citep[6dFGRS;][]{Jones2006}, and the Sloan Digital Sky Survey Data Release 7 \citep[SDSS;][]{Abazajian2009}. This dataset contains approximately 69,000 galaxies, extending to redshifts of $z < 0.1$ \citep[see figure 2 in][ for a visualization of the catalogue's coverage and geometry]{Jasche2019}.

The input \tmpp galaxies are categorized into eight equally spaced bins of $K$-band absolute magnitude spanning the range $-25 \leq M_{\mathrm{K}} < -21$. To account for the two distinct flux completeness limits in the \tmpp catalogue, we further divide the galaxies into two apparent magnitude intervals: $8 \leq m_{\mathrm{K}} < 11.5$ and $11.5 \leq m_{\mathrm{K}} < 12.5$. To capture any potential redshift evolution in the bias parameters, we additionally split the sample into two redshift bins: $0 < z < 0.03$ and $0.03 < z < 0.1$. The choice of $z = 0.03$ as the dividing line reflects the redshift extent of a smaller companion inference, dubbed \manticoremini (see \cref{sect:inference_setup,sect:borg_performance}), designed to enable a consistent comparison between \manticorelocal and \manticoremini across overlapping volumes.

These selection cuts effectively impose both bright and faint limits on the data, beyond which the inference remains largely unconstrained due to the absence of observational information. Consequently, the region meaningfully constrained by the data spans approximately 12 to 350~Mpc in distance (corresponding to $z \approx 0.004-0.08$ for $H_0 = 68.1$~km~s$^{-1}$~Mpc$^{-1}$). However, the signal-to-noise ratio drops significantly beyond $r \sim 200$~Mpc, motivating our definition of a fiducial constrained region within $r < 200$~Mpc (or $z \lesssim 0.05$, as demonstrated in \cref{sect:borg_performance}).

The catalogue's angular completeness is characterized by a high-resolution, two-dimensional mask \citep[see figure 4 in][]{Lavaux2011}, which encodes the ratio of spectroscopic redshifts to photometric targets across the sky. This mask accounts for survey incompleteness due to factors such as bright foreground sources and Galactic extinction, as well as redshift completeness, which is incorporated into the \borg forward modelling framework. To account for the survey’s flux limit, we apply a radial selection function derived from the Schechter luminosity function \citep{Schechter1976}, which reconstructs the expected absolute magnitude distribution for a given apparent magnitude and redshift. The radial selection function is constructed using the $K$-band Schechter luminosity function parameters from \citet{Lavaux2011}, with $\alpha = -0.94$ and $M_* - 5 \log_{10} h = -23.28$.  

Together, the survey’s angular and radial selection effects define the survey’s response operator (denoted $R$ in \cref{sect:borg_algorithm}), a three-dimensional selection function integrated into the \borg forward model to accurately emulate observational constraints \citep[see][for further details]{Jasche2013,Jasche2015}.

\subsection{Inference setup for the \borg algorithm}
\label{sect:inference_setup}

\begin{table}
    \centering
    \begin{tabular}{lccc} 
        \hline
        Inference & $L$ & $L_{\mathrm{WNF}}$ & $R_{\mathrm{constrained}}$ \\ 
        - & [Mpc] & [Mpc] & [Mpc] \\
        \hline
        \textbf{\manticorelocal} & \textbf{1000} & \textbf{3.9} & $\approx$ \textbf{200} \\  
        \manticorelocallo & 1000 & 7.8 & $\approx 200$ \\  
        \manticoremini & 500 & 3.9 & $\approx 125$ \\  
        \hline
    \end{tabular}
    \caption{Summary of the \borg field-level-inference configurations produced by this study. From left to right; the size of the parent domain, the inference grid resolution (indicated by comoving cell size), and the radial extent of the region constrained by data. Note that \manticorelocal is the fiducial model for this study, which forms the basis for all the presented results.}
    \label{table:manticore_products}
\end{table}

To infer the posterior distribution of initial conditions, we employ a modified version of the \borg algorithm, referred to as the \manticore model. The \borg framework \citep{Jasche2019} performs full field-level Bayesian inference of the large-scale structure by forward-modelling the gravitational evolution of the Universe from Gaussian initial conditions to the observed galaxy distribution, while accounting for observational effects such as selection functions and survey geometry. This is achieved via Hamiltonian Monte Carlo sampling in a high-dimensional space of initial density fields, jointly inferring structure formation histories, galaxy bias parameters, and the underlying matter distribution. 

Building on this foundation, our \manticore model incorporates several key improvements: a more flexible galaxy bias model with enforced ergodicity, a suite of physics-informed priors designed to preserve statistical consistency with \lcdm expectations, and a generalized Poisson likelihood that allows for the over-dispersion of galaxy counts. A detailed overview of the original \borg algorithm and the specific modifications introduced in \manticore can be found in Appendix~\ref{sect:borg_algorithm}.

For this work, we follow the inference methodologies established in previous Aquila Consortium studies \citep[e.g.,][]{Jasche2019,Stopyra2024_COLA}, operating within a parent domain of side length $L = 1000$~Mpc, with the fiducial observer positioned at the centre. The inference is performed on a Cartesian equidistant grid with $N = 256^3$ elements, providing a spatial resolution of approximately $3.9$~Mpc in the phases of the initial white noise field.  

 The 32 galaxy subcatalogues (defined in \cref{sect:2mpp}) each have their own independent set of initially unknown galaxy bias parameters. These are inferred jointly with the phases of the white noise field during the analysis (see \cref{sect:borg_algorithm} for details).

The inference adopts cosmological parameters from the Dark Energy Survey year three \citep[DES Y3,][]{DEScosmo} '3 × 2pt + All Ext.' \lcdm cosmology: $h=0.681$, $\Omega_\mathrm{m} = 0.306$ , $\Omega_\mathrm{\Lambda} = 0.694$, $\Omega_\mathrm{b} = 0.0486$, $A_\mathrm{s} = 2.099 \times 10^{-9}$, $n_\mathrm{s} = 0.967$, $\sigma_\mathrm{8} = 0.807$. This choice aligns closely with the fiducial cosmology of the state-of-the-art \flamingo simulation suite \citep{Schaye2023}.

This configuration defines our fiducial inference setup, referred to as \squotes{\manticorelocal}. In addition to this baseline run, we produce two supplementary inference products: \manticorelocallo, a lower-resolution version of the fiducial volume, and \manticoremini, an inference at the same resolution as \manticorelocal conducted within a smaller parent domain. While these configurations differ in domain size, grid resolution, and the extent of the data-constrained region, all other aspects of the inference setup—including the selection function, bias model, and likelihood—remain identical. A summary of the three configurations is provided in \cref{table:manticore_products}.

These alternative runs are used primarily for convergence testing, assessing the robustness of inferred cluster properties and radial density profiles to changes in resolution and volume size, for example see \cref{sect:resolution_convergence}. Additionally, the smaller parent volume with its more limited observational constraints may, in future work, help identify which specific features of the data influence the formation of particular structures.

All analyses presented in this work are based on the inferred posterior and subsequent resimulations of the fiducial \manticorelocal configuration.

\subsection{Posterior resimulations of the initial conditions}
\label{sect:posterior_resimulations}

Using the inferred initial conditions, we generate fifty dark-matter-only (DMO) \textit{posterior resimulations} of the local Universe. Each resimulation begins with an independent sample drawn from the \borg posterior of white noise fields, from which particle initial conditions are generated according to the corresponding cosmology. These initial conditions are then evolved forward to $z=0$ using a high-accuracy $N$-body solver. This procedure acts as a strict posterior predictive test: rather than relying on the output of the inference directly—which uses an approximate \texttt{COLA} gravity model—we resimulate the inferred initial conditions with a full $N$-body code to achieve higher physical fidelity. This methodology has been applied in previous generations of Aquila works, including the \texttt{CSiBORG} resimulations \citep{2021PhRvD.103b3523B,Stiskalek2025} based on the initial conditions of \citet{Jasche2019}, and more recently in the work of \citet{Stopyra2024_COLA}.

The particle initial conditions are constructed using the \texttt{MONOFONIC} algorithm \citep{Hahn2020, Michaux2021}. The constrained white noise fields inferred by \borg are defined on a $256^3$ grid within a $L=1000$~Mpc domain, corresponding to a spatial resolution of 3.9~Mpc. To enable higher-resolution $N$-body simulations, we oversample this grid by placing $1024^3$ dark matter particles, equating to a particle mass of $3.7 \times 10^{10}$~\msol. 

As the inference is only defined up to the Nyquist frequency of the $256^3$ grid, \texttt{MONOFONIC} augments the initial conditions by injecting Gaussian random modes on smaller scales, beyond the resolution of the constraints. This approach ensures that the large-scale structure is consistent with the Bayesian inference, while the small-scale power is drawn from an unconstrained realization consistent with the assumed cosmology. The final initial conditions are generated at $z = 69$ using second-order Lagrangian perturbation theory.

We note that while the large-scale structure of each realization is determined by the inferred constrained modes from \borg, the small-scale Fourier modes—beyond the Nyquist limit of the inference grid—are randomly sampled. Consequently, the posterior variance across our resimulations reflects uncertainty from both constrained large-scale modes and stochastic small-scale completions. Fully disentangling these contributions would require sampling many small-scale realizations per posterior sample—a computationally prohibitive procedure at our resolution, but a valuable direction for future work \citep[see e.g.,][]{Sawala2021a}.

The subsequent evolution to $z=0$ is carried out with \texttt{SWIFT} \citep{Schaller2024}, a high-precision, open-source simulation code for gravity, hydrodynamics, and galaxy formation\footnote{Publicly available at \href{http://www.swiftsim.com}{www.swiftsim.com}.}. The resimulations adopt the same cosmological parameters used in the forward model during inference, ensuring direct comparability with the inferred posterior distribution.

For our analysis, we use the $z=0$ particle snapshot rather than a lightcone, assuming negligible evolution in the positions and properties of haloes within the constrained volume ($0 < z < 0.045$). This choice simplifies direct comparisons with observational catalogs while preserving the statistical integrity of the reconstructed structures.

This suite of DMO resimulations, referred to as the \manticorelocal posterior resimulations, serves as the foundation for the results presented in \cref{sect:results}.

\subsubsection{Structure finding}

Haloes and substructures in the $z=0$ outputs of the posterior resimulations are identified using the \texttt{HBT+} (\textit{Hierarchical Bound-Tracing}) subhalo finder\footnote{Publicly available at \href{https://github.com/Kambrian/HBTplus}{https://github.com/Kambrian/HBTplus}.} \citep{Han2012, Han2018}.

A key comparison in this study involves the local cluster population, requiring a consistent definition of total halo mass. We adopt \masscrit, the total mass enclosed within \rcrit, the radius within which the mean enclosed density is 200 times the critical density of the Universe ($200 \times \rho_{\mathrm{crit}}$). Where necessary, literature estimates quoted as $M_{\mathrm{500,crit}}$ are converted to \masscrit using the concentration-mass relation from \citet{Bhattacharya2013}.

Structures with at least 20 bound dark matter particles are retained, corresponding to $\texttt{Mbound} \geq 7.3 \times 10^{11}$~\msol, though for this work we only consider well resolved central haloes with masses above \masscrit $\geq 10^{13}$~\msol ($> 200$ particles). The position and velocity of haloes are inferred from their most bound dark matter particle, i.e., $\texttt{ComovingMostBoundPosition}$ and $\texttt{PhysicalMostBoundVelocity}$.

\subsubsection{Control simulations}

To evaluate whether the data-constrained regions in the posterior resimulations resemble a \lcdm Universe, we require a suitable random \lcdm simulation suite as a control.

We therefore additionally generate ten control simulations with the same domain size, resolution, and cosmology as the posterior resimulations. These follow the same pipeline used to generate the posterior resimulations, but are seeded with random, unconstrained phases. Collectively, they form our control sample and are denoted as $\texttt{Random-$\Lambda$CDM}$ in the figures.
\section{Results}
\label{sect:results}

In this section, we present the results of the \manticorelocal posterior resimulations. 
We begin in \cref{sect:lcdm_parent} by assessing their statistical consistency with \lcdm expectations. 
Next, in \cref{sect:local_supervolume}, we examine the properties of the reconstructed density and velocity fields within the local supervolume (\(R < 200\)~Mpc). 
Finally, in \cref{sect:cluster_masses} we compare the inferred local cluster population to observational data.

\subsection{A \lcdm-like parent volume}
\label{sect:lcdm_parent}

\begin{figure}
    \includegraphics[width=\columnwidth]{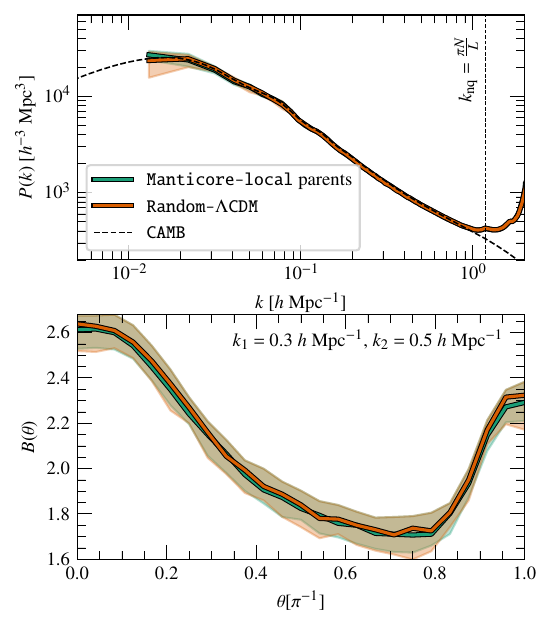}
    \caption{Matter power spectrum ($\textit{upper panel}$) and bispectrum ($\textit{lower panel}$) at $z = 0$ for the posterior resimulations parent volumes, compared to the control suite of simulations and the $\texttt{CAMB}$ theoretical prediction. Solid lines indicate median values, with shaded regions representing the \tentoninety percentile range. The posterior resimulations show excellent agreement with the control simulations across all scales, supporting consistency with \lcdm expectations in both linear and nonlinear regimes.}
    \label{fig:parent_power_spectrum}
\end{figure}

The \lcdm framework provides the most successful description of the large-scale structure and evolution of the Universe. Given its empirical success, we choose to explicitly test whether the observed data of the local Universe can be well described within the this framework. To do so, we perform a series of post-inference consistency checks to assess whether the parent volumes (i.e., the full $L=1000$~Mpc domain) of the posterior resimulations exhibit statistical properties consistent with \lcdm expectations. Other than the Gaussianity of the initial white noise field, these evaluations were not considered during the inference itself, but instead serve to validate the physical plausibility and statistical consistency of our inferred realizations.

\begin{itemize}

    \item \textit{Gaussianity of the Initial White Noise Field}: 
    The \lcdm paradigm assumes that the initial density perturbations form a Gaussian random field, which serves as the foundation for linear structure formation. Any deviation from Gaussianity in the reconstructed white noise fields could indicate systematic issues in the inference process, or inconsistencies with the assumed model. We test this by evaluating the power spectrum of the white noise field, its probability density function (PDF), and a Quantile-Quantile (Q-Q) comparison assessing the phase distribution relative to the theoretical normal distribution.

    \item \textit{Halo Mass Function (HMF)}: 
    The halo mass function provides a well-characterized prediction of structure formation in \lcdm cosmology. We compare the HMF of the reconstructed parent volumes at \( z = 0 \) to that of an ensemble of control \lcdm simulations with the same cosmological parameters. Agreement between these distributions ensures that the growth of structure in the inferred volumes follows \lcdm expectations across the full mass range of resolved haloes.

    \item \textit{Power Spectrum and Bispectrum}: 
    The power spectrum encodes the clustering of matter across scales, while the bispectrum captures non-Gaussian features introduced by nonlinear structure formation. Agreement between the power spectrum and bispectrum of the reconstructed density field at $z=0$ and those of control \lcdm simulations confirms that both linear and nonlinear aspects of structure formation are correctly reproduced in our parent volumes.

\end{itemize}

As shown in \cref{fig:parent_power_spectrum}, the posterior ensemble matches the \lcdm theoretical prediction and control simulations in both power spectrum and bispectrum. Additional validation plots, including the halo mass function and white noise field statistics, are provided in Appendix~\ref{sect:parent_lcdm_cont}, where we find that all tests confirm the statistical consistency of our inferred volumes with the \lcdm model.

\subsection{The matter field of the local supervolume}
\label{sect:local_supervolume}

With the reconstructed parent volumes from the posterior resimulations validated as \lcdm-like realizations, we now turn to examining the statistical properties of the constrained region within them. Specifically, we focus on the inner 200 Mpc surrounding the fiducial observer, referred to as the local supervolume \citep{Stopyra2021}. This region encompasses the volume where observational data constraints are strongest, making it an ideal testbed for assessing how well the inferred structures align with expectations within the \lcdm framework.

Our analysis takes a primarily theoretical approach, providing predictions for the local supervolume's statistical properties. We examine the spatial correspondence between the reconstructed density field and observed galaxy distribution, followed by analyses of the halo mass function, radial density profile, and abundance of massive clusters. Together, these diagnostics offer insights into the local Universe's density environment and its role in cosmic structure formation.

\subsubsection{The galaxy distribution and the inferred dark matter field}
\label{sect:visual_lss}

\begin{figure}
    \includegraphics[width=\columnwidth]{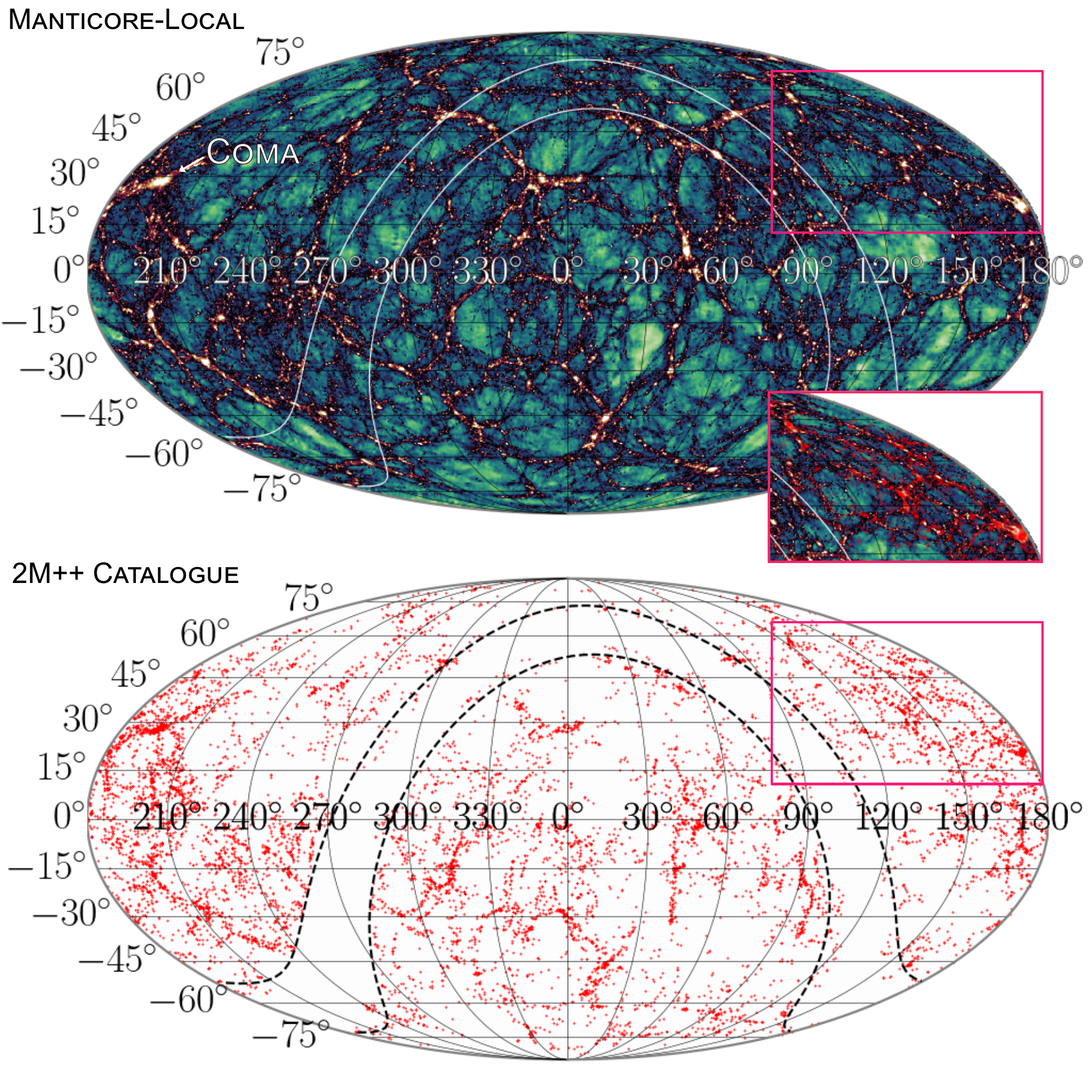}
    \caption{Spatial distribution of reconstructed dark matter density and observed galaxy positions in the local supervolume. The upper panel shows a Mollweide projection in equatorial coordinates of the dark matter density field from a single \manticorelocal posterior realization, for a shell spanning 95–125 Mpc from the observer. The lower panel displays the galaxy distribution from the \tmpp catalogue within the same shell. In both panels, the Zone of Avoidance is overlaid for reference. The inset panel presents a highlighted view of a representative sky region, overlaying the galaxy positions directly on the reconstructed density field, illustrating the close correspondence between observed galaxies and inferred large-scale structure.}
    \label{fig:2mpp_with_dm}
\end{figure}

\begin{figure}
    \includegraphics[width=\columnwidth]{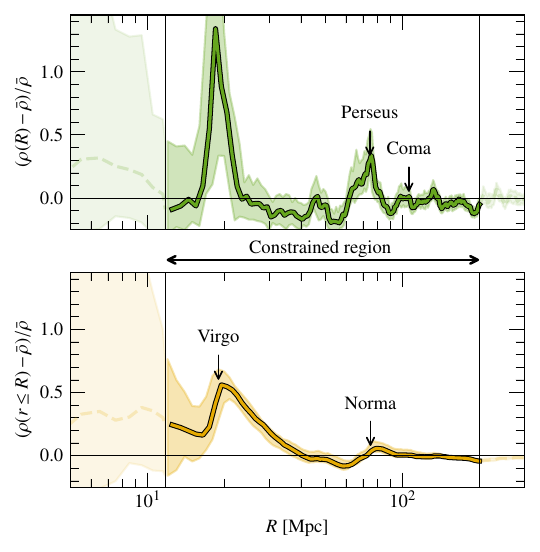}
    \caption{Radial density profiles of the local supervolume relative to the cosmic mean density. The upper panel shows density in spherical shells, revealing distinct features corresponding to major structures including the prominent Virgo Cluster ($\approx$16 Mpc), and the Perseus-Coma complex ($\approx$75-100 Mpc). The lower panel presents density in cumulative spheres, showing how local overdensities integrate into the larger cosmic context. Solid lines represent median values across posterior resimulations, with shaded regions indicating the \tentoninety percentile range. The constrained region (12-200 Mpc, see \cref{sect:borg_performance} for definition details) is highlighted, with faded areas representing volumes where our reconstruction provides less reliable predictions.}
    \label{fig:radial_density_profile}
\end{figure}

\begin{figure}
    \includegraphics[width=\columnwidth]{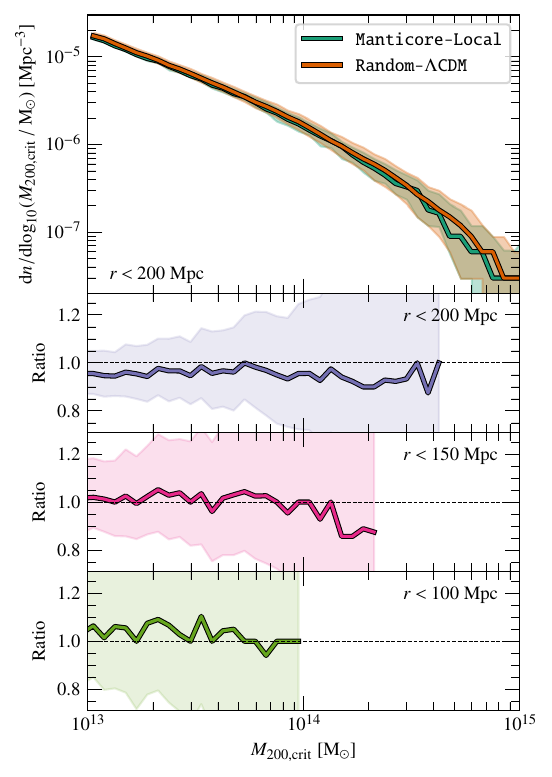}
    \caption{Halo mass function within the local supervolume. The upper panel shows the median mass function within $R=200$ Mpc with its 10-90 percentile uncertainty range from posterior resimulations, compared with measurements from randomly placed spheres of equivalent volume in our \lcdm control simulations. The lower panels display the ratio between the reconstructed and control mass functions at three different scales: $R<200$ Mpc, $R<150$ Mpc, and $R<100$ Mpc. This multi-scale analysis reveals a systematic variation with radius: approximately 5\% fewer haloes than cosmic mean expectations at $R<200$ Mpc, transitioning to approximately 5\% more haloes at $R<100$ Mpc. This radial dependence demonstrates the importance of accounting for cosmic variance when interpreting observations with different depth limitations.}
    \label{fig:inner_halo_mass_function}
\end{figure}

The cosmic web—composed of an intricate network of filaments, clusters, walls, and voids—constitutes the fundamental large-scale structure of the Universe \citep{Bond1996, Colberg2005}. This complex architecture emerges through the gravitational collapse of primordial density fluctuations, directing matter flow along filaments toward high-density nodes, ultimately forming galaxy clusters. Analyzing the statistical properties of these structures yields essential constraints for cosmological models \citep{Klypin2011, Cautun2014}.

\cref{fig:2mpp_with_dm} visually compares a realization of the posterior dark matter density field with the observed galaxy distribution from the \tmpp catalogue. This visualization, encompassing a radial slice spanning distances of 95–125 Mpc from the observer, highlights prominent structures such as the Coma Cluster, clearly illustrating the spatial correlation between inferred matter densities and observed galaxies. The evident correspondence between density peaks in the reconstruction and galaxy positions affirms that our Bayesian inference successfully integrates observational constraints into dynamically coherent cosmological realizations. Additionally, the visualization demonstrates that regions lacking direct observational constraints—such as the Zone of Avoidance—are plausibly reconstructed according to the prior expectations of the \lcdm cosmological model.

While the visual agreement is compelling, a quantitative cross-correlation between density and galaxy fields is non-trivial due to the nonlinear and scale-dependent nature of galaxy bias. Robust interpretation requires matching to mock galaxy catalogs generated from our initial conditions—something only possible with hydrodynamic simulations, which we leave to future work.

Having established this visual correspondence, we now proceed to statistical analyses of the local supervolume, quantifying its properties through critical cosmological metrics.

\subsubsection{Radial density profile}
\label{sect:radial_density_profile}

The radial density profile provides a fundamental diagnostic of the matter distribution in our cosmic vicinity, probing both the transition to large-scale homogeneity and the possible presence of local anomalies. By measuring the enclosed matter density as a function of distance from the observer, this profile directly tests the cosmological principle at intermediate scales \citep{Scrimgeour2012} and informs ongoing discussions around features such as the so-called “local hole” or “KBC void” \citep{Keenan2013, 2020MNRAS.499.2845H}.

We assess the radial distribution of matter using two complementary statistics: the density within concentric spherical shells and the cumulative density enclosed within spheres of increasing radius. Both measures are normalized by the mean cosmic matter density and presented in \cref{fig:radial_density_profile}. We highlight the constrained region (approximately 12–200~Mpc), defined as the contiguous volume where the inferred signal dominates over the noise (see \cref{sect:borg_performance}). Outside this region, the reconstructions become progressively less predictive.

The shell-based profile, presented in the upper panel, reveals sharp density enhancements at distances corresponding to known structures. A strong overdensity appears within the central 20~Mpc, driven by the Virgo cluster, while additional peaks are evident at $\approx 75$–100~Mpc, coinciding with the Coma and Perseus clusters.

The cumulative density profile offers a smoothed view, integrating over local fluctuations to highlight broader density trends. Beyond the Virgo region, the profile remains largely stable, with density deviations typically within $\pm5$\% of the cosmic mean throughout the range $50 < R < 200 $~Mpc. This behaviour is broadly consistent with expectations from \lcdm cosmology, which predicts a transition to homogeneity on scales of a few hundred Mpc \citep{Wu2017}.

These results provide context for observational claims of a large-scale local underdensity—most notably, the “KBC void” proposed by \citet{Keenan2013} and further analysed by \citet{2020MNRAS.499.2845H}. Their work suggests a significant underdensity ($\delta \sim 0.5$) extending from $R \sim 40$ to at least 300~Mpc, with recovery to the mean density only occurring near $R \sim 500$~Mpc. Such a deep and extended void is argued to be highly unlikely in \lcdm (with a statistical tension exceeding $6\sigma$ in some analyses), and has been proposed as a potential contributor to the observed Hubble tension.

By contrast, our constrained reconstructions find no evidence for a large-scale underdensity of that magnitude within the region $R < 200$~Mpc. While mild fluctuations are present, they remain within the expected cosmic variance and do not indicate a coherent void structure of the required depth. However, important caveats must be considered. First, our analysis is explicitly limited to a constrained volume of radius 200~Mpc. If the proposed underdensity extends significantly beyond this scale, as claimed, our inference may lack the spatial coverage necessary to detect it.

Second, the constrained region is embedded within a larger periodic simulation volume of size $L = 1000$~Mpc, which, by construction, enforces global density neutrality. This implies that any large-scale underdensity in the central region must be compensated by overdensities elsewhere, potentially limiting the capacity of our model to represent extreme deviations from the mean density. Whether this introduces a significant bias remains unclear, but it does emphasize the importance of scale when interpreting constrained density profiles.

To robustly evaluate the presence of extended underdensities, future constrained realizations should ideally combine larger parent volumes with observational data that reaches to greater depths. This would allow the inner region to vary more freely, while still maintaining statistical consistency over the full simulation domain. Our current results suggest that the Local Universe, within $R < 200$~Mpc, is broadly consistent with \lcdm expectations in terms of mean density and its radial behaviour—but a definitive assessment of more extended void scenarios will require additional data and modelling.

\subsubsection{The halo mass function}
\label{sect:inner_halo_mass_function}

Standard simulations produce halo mass functions representative of the cosmic average, which has become the default expectation in cosmological analyses. However, the local Universe as observed from our location is neither periodic nor necessarily at the mean density, introducing an important source of cosmic variance. The unique properties of the local supervolume provide a distinct window into cosmic structure formation that differs fundamentally from analyses based on idealized cosmological volumes. The impact of this local variance is particularly relevant for shallow galaxy surveys, such as the upcoming $\textit{Widefield ASKAP L-band Legacy All-sky Blind surveY}$ (WALLABY) survey \citep{Koribalski2020}, whose selection function and completeness will be intimately tied to the underlying dark matter distribution specific to our cosmic neighbourhood. This issue also extends to deeper surveys such as the $\textit{4-metre Multi-Object Spectroscopic Telescope}$ (4MOST) Hemisphere Survey \citep{2019Msngr.175....3D}, which aims to map the large-scale velocity field out to distances of $\approx 500$ Mpc and may likewise be influenced by the density fluctuations characterizing our local environment.

\cref{fig:inner_halo_mass_function} presents the median halo mass function of the local supervolume derived from our posterior resimulations, along with its associated uncertainty range (\tentoninety percentiles), compared against measurements from 500 random spheres of equivalent volume drawn from our control simulations. The lower panels of this figure display the halo mass function ratios for three different radial cuts: $R<200$ Mpc (i.e., the ratio of the two lines in the first panel), $R<150$ Mpc, and $R<100$ Mpc. This multi-scale analysis reveals a systematic variation in halo abundance that depends critically on the volume considered.

At the largest scale ($R<200$ Mpc), we observe approximately 5\% fewer haloes than predicted by \lcdm \citep[similar to previous Aquila works, e.g.,][]{2022MNRAS.512.5823M,Stopyra2024_COLA}. This deficit aligns with our previous measurements showing the local environment to be underdense by $\approx 5\%$ at equivalent scales (as shown in the lower panel of \cref{fig:radial_density_profile}). However, as we focus on progressively smaller volumes, this trend reverses. Within $R<100$ Mpc, we find approximately 5\% more haloes than the cosmic mean would suggest, reflecting the enhanced density associated with prominent nearby structures like Virgo, Perseus, and Coma. The direct relationship between environmental density and halo abundance is a fundamental prediction of both linear perturbation theory and hierarchical structure formation in \lcdm \citep{Press1974,Sheth1999}.

The observed radial dependence of halo abundance underscores the critical need to account for cosmic variance when interpreting local Universe observations. Surveys with varying depth limitations effectively sample different cosmic environments, which may introduce systematic biases in derived cosmological parameters if this variance remains unmodelled. For now, the significant scatter across our different \manticore realizations ($\approx$10-20\%) reveals the inherent uncertainty in precisely constraining both the local density and corresponding halo abundance within $R<200$~Mpc. This highlights the considerable value of future deep, wide-field surveys of the local Universe (out to $z \sim 0.1$), which would provide tighter constraints on both the amplitude and scale-dependence of local density fluctuations, thereby significantly reducing systematic uncertainties in cosmological inferences derived from nearby structures.

\subsubsection{The abundance of massive clusters}
\label{sect:abundance_of_massive_clusters}

\begin{figure}
    \includegraphics[width=\columnwidth]{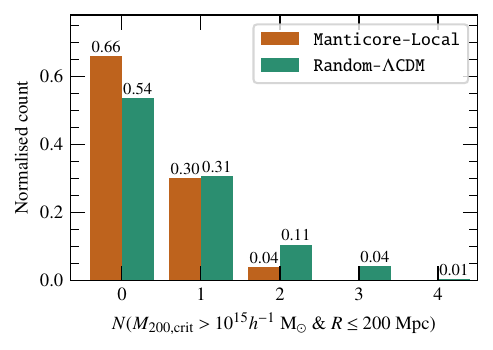}
    \caption{The fraction of realizations containing \( N \) clusters with masses exceeding \( 10^{15} \)~\msolh within 200~Mpc of the observer. The figure shows the distribution from the \manticorelocal posterior resimulations and the distribution from 500 randomly placed $R=200$~Mpc spheres in the \lcdm control simulations. Over 85\% of posterior realizations contain either zero or one such massive cluster, which lies within the expected range from \lcdm. Small differences between the distributions are not unexpected: while the \lcdm controls explore the full range of outcomes allowed by cosmic variance, the \manticorelocal simulations are constrained by observational data, and therefore tend to favour a specific realization of the local Universe.}
    \label{fig:N_above_15}
\end{figure}

The abundance of the most massive haloes, particularly those exceeding $10^{15}$~\msolh ($1.47 \times 10 ^{15}$~\msol for $H_0 = 68.1$~km~s$^{-1}$~Mpc$^{-1}$), provides a sensitive test of the \lcdm framework. The steep exponential drop in the halo mass function at these scales means that even small deviations in the number of such clusters can signal potential tensions with cosmological expectations \citep{Press1974, Tinker2008}. The test is highly sensitive to cosmological parameters, especially \( \sigma_8 \): increasing $\sigma_8$ boosts the expected number of rare massive haloes, while recent weak lensing results favouring lower values (e.g. \citealt{Heymans2021}) would predict fewer such objects \citep{Kim2015, Bocquet2016}.

\citet{Stopyra2021} investigated this idea using the local supervolume ($R < 200$~Mpc), and found that the number of observed clusters above $10^{15}$~\msolh varies from 0 to 5 depending on the chosen mass estimator. These choices led to dramatically different \lcdm likelihoods — from order unity to values as low as $10^{-5}$. The study showed that even seemingly modest changes in estimated cluster masses can significantly affect any inferred tension with \lcdm. However, it also highlighted the challenge: mass estimates from X-ray/SZ observations, weak lensing, and dynamical methods often disagree by factors of several, complicating robust cosmological interpretation.

In \cref{fig:N_above_15}, we compare the predicted number of $M > 10^{15}$~\msolh clusters within 200~Mpc in our \manticorelocal posterior resimulations against the control suite of \lcdm simulations sampled with randomly placed spheres. In over 85\% of constrained realizations, either zero or one such cluster is found—matching the preferred range from the \lcdm control simulations. However, as our simulations are conditioned on observed galaxy data, they are naturally biased toward a specific realization of the Universe. The control sample, by contrast, reflects the full cosmic variance permitted under \lcdm.

This distinction is crucial: while the control simulations provide a statistical baseline, the constrained \manticorelocal realizations reflect the specific structure of the actual local Universe. Thus, the lower abundance of massive haloes inferred from \manticorelocal is not necessarily reflecting any tension with \lcdm, but instead highlights how real-world constraints modulate structure formation \citep[see also][]{2022MNRAS.516.3592H}. For example, the boosted preference of having zero or one such massive cluster within 200 Mpc, relative to the \lcdm expectation, could be linked to the reported underdensity at this scale (see \cref{fig:radial_density_profile}).

Compared to earlier analyses that relied on simplified likelihood modelling and cluster mass estimates derived from a mixture of observational methods and assumptions \citep{2011JCAP...07..004H,2013JCAP...07..022H,Stopyra2021}, our approach leverages a fully dynamical forward model, robust Bayesian inference, and a well-validated nonlinear bias prescription. In that sense, it delivers a more physically motivated and statistically coherent test of local halo abundance. Future improvements in mass calibration—particularly via weak lensing or hydro-dynamical forward modelling—will be critical to further sharpen this diagnostic and resolve lingering discrepancies in observed cluster mass estimates.

Overall, our results indicate that the local supervolume, as inferred by \manticorelocal, is not anomalous in its massive halo content, and lies well within the range of \lcdm expectations.

\subsection{The velocity field of the local supervolume}
\label{sect:velocity_field}

\begin{figure}
    \includegraphics[width=\columnwidth]{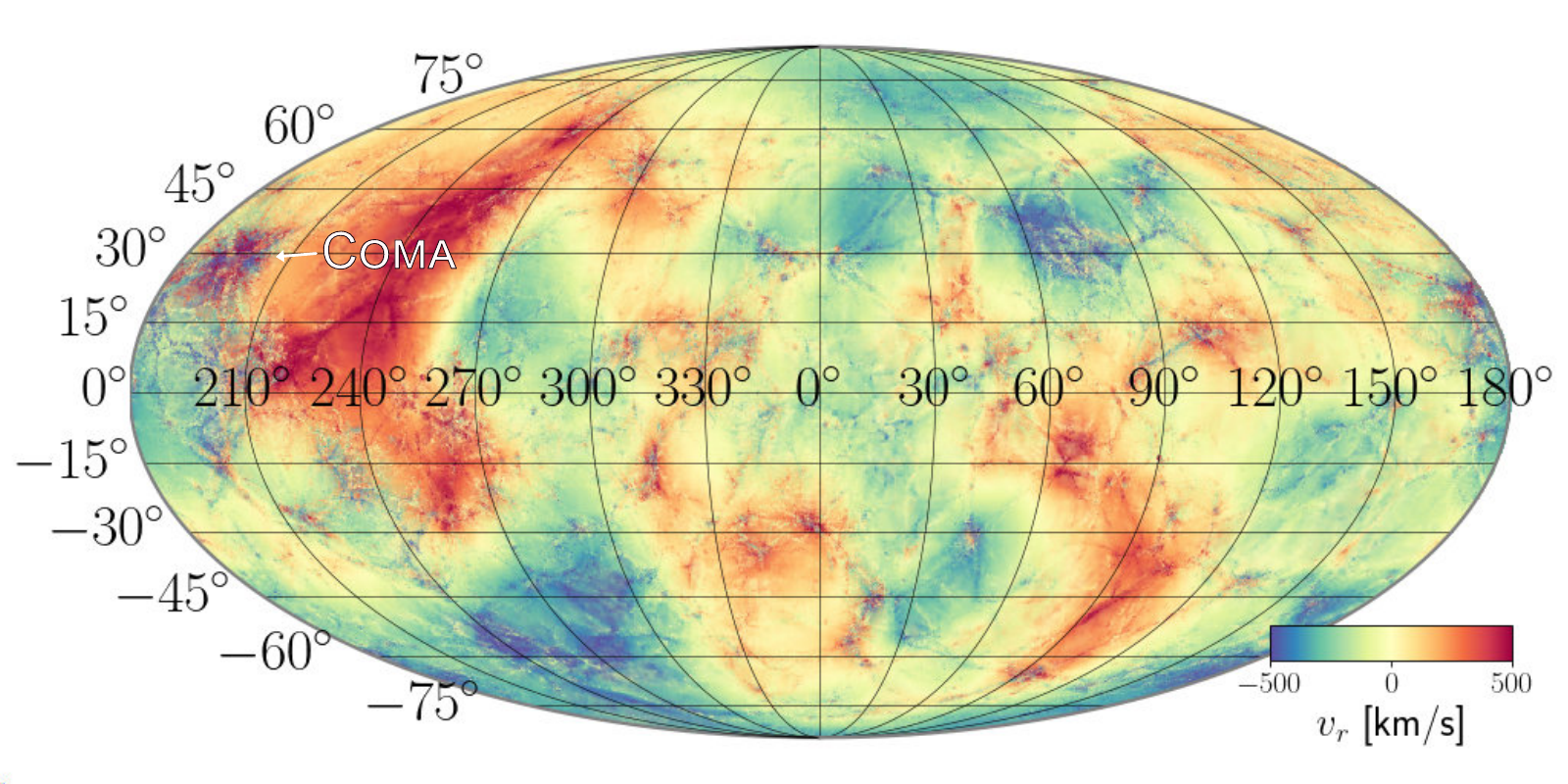}
    \caption{Mollweide projection of the radial peculiar velocity field from a single \manticorelocal posterior realization in a shell spanning 95 to 125~Mpc. Blue regions correspond to coherent infall (motion toward the observer), while red indicates outflow. The map captures the complex velocity structure associated with large-scale gravitational potentials, such as the Coma cluster.}
    \label{fig:vr_field}
\end{figure}

Peculiar velocities—deviations from the uniform Hubble flow—trace the gravitational accelerations induced by the surrounding matter distribution \citep{1983ApJ...267..465D,1995PhR...261..271S}. They provide a direct probe of the total matter field, including both visible and dark components, and offer a powerful complement to galaxy distributions in studies of cosmic structure. The peculiar velocity field plays a central role in cosmology: it enables corrections to redshift-independent distance indicators, constrains the growth rate of structure (\(f\sigma_8\)), and tests \lcdm predictions through bulk flow and velocity correlation analyses \citep{Yahil1991, Pike2005, Davis2011, Watkins2009, Riess2011, Boruah2020, 2024MNRAS.527.3788H,2024arXiv241119484T,Stiskalek2025}.

The \borg framework provides a self-consistent prediction for the local peculiar velocity field, one that is indirectly inferred via the field-level Bayesian reconstruction constrained by galaxy observations. Unlike linear theory reconstructions, \borg captures fully non-linear dynamics through forward modelling of the gravitational evolution from initial conditions. \Cref{fig:vr_field} shows a radial projection of the velocity field from a single \manticorelocal posterior realization in the 95–125~Mpc shell, corresponding to the same region displayed in \cref{fig:2mpp_with_dm}. The map reveals complex flows around massive structures, such as the Coma cluster, as well as large-scale coherent motions. In observationally obscured regions, such as the Zone of Avoidance, the inferred flows remain physically plausible under the prior expectations of the \lcdm model.

To quantitatively assess the accuracy of our inferred velocity field, we adopt the Bayesian model comparison framework introduced by \citet{Stiskalek2025}. This approach evaluates how well reconstructed velocity fields explain independent observational datasets, including five major peculiar velocity catalogs: 2MTF \citep{2008AJ....135.1738M,2019MNRAS.487.2061H,Boruah2020}, SFI++ \citep{2006ApJ...653..861M,Boruah2020}, CF4 Tully-Fisher samples \citep{2023ApJ...944...94T}, and the LOSS \citep{2011MNRAS.412.1441L,2013MNRAS.433.2240G} and Foundation \citep{2018MNRAS.475..193F} supernova samples. The Bayesian evidence calculation marginalizes over key nuisance parameters, such as external flow (\( V_{\mathrm{ext}} \)), and small-scale velocity dispersion (\( \sigma_v \)), yielding a robust and interpretable comparison of model fidelity.

\begin{figure}
    \includegraphics[width=\columnwidth]{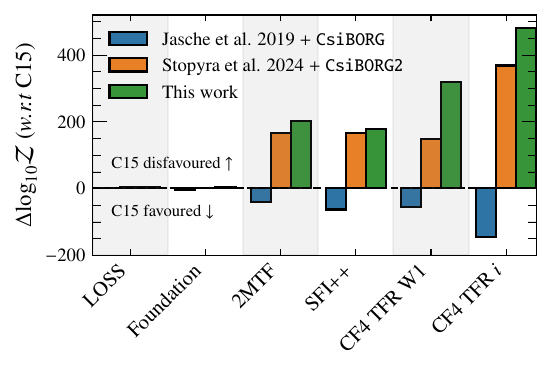}
    \caption{Bayesian evidence comparison for various velocity field reconstructions, evaluated across five independent empirical peculiar velocity datasets, using the framework of \citet{Stiskalek2025}. Bars show relative differences in $\log_{10}$ evidence with respect to the \citet{Carrick2015} linear theory model. \manticorelocal (this work) consistently achieves the highest evidence across all datasets, indicating the best overall agreement with observed galaxy peculiar velocities.}
    \label{fig:velocity_evidence_compare}
\end{figure}

\Cref{fig:velocity_evidence_compare} presents the outcome of this comparison. A clear progression in performance is observed across successive reconstruction efforts by the Aquila consortium. The earlier \borg-based initial conditions from \citet{Jasche2019} were the foundation for the high resolution \texttt{CSiBORG} suite of resimulations \citep{2021PhRvD.103b3523B}. These were followed by the \texttt{CSiBORG2} resimulations~\citep{Stiskalek2025}, built upon the initial conditions from \citet{Stopyra2024_COLA}, which extended the \borg framework to a \texttt{COLA}-enhanced forward model. Both of these high-resolution velocity fields, and others, were benchmarked by \citet{Stiskalek2025} in their so-called ``Velocity Field Olympics,'' a systematic comparison of constrained reconstructions based on a range of methodologies—including linear theory models \citep{Carrick2015}, Wiener-filter-based reconstructions \citep[][]{Sorce2018}, linear forward modeling approaches from peculiar velocity data \citep[][]{2023A&A...670L..15C}, and machine learning reconstructions \citep{2024A&A...689A.226L}.

Among all models tested in that study, \texttt{CSiBORG2} emerged as the highest-performing reconstruction across the velocity datasets. The \manticorelocal posterior resimulations now surpass even \texttt{CSiBORG2} on all five independent catalogs, achieving the highest Bayesian evidence in every case. Note that the magnitude of the Bayesian evidence is naturally sensitive to the size of each dataset, with catalogs containing more velocity tracers (e.g. CF4 TFR) yielding larger absolute differences in $\log_{10}$ evidence. 

The consistent preference for \manticorelocal across a diverse set of tracers—spanning supernovae, Tully–Fisher distances, and varying redshifts and sky coverage—underscores the robustness and accuracy of the reconstructed velocity field. By embedding observational constraints within a physically motivated dynamical model, \manticorelocal sets a new benchmark for the constrained modelling of local velocity flows, with broad implications for studies of bulk motions, the Hubble tension, and kinetic Sunyaev–Zel’dovich effects in the nearby Universe.

\subsection{The cluster population of the local Universe}
\label{sect:cluster_population}

Galaxy clusters are the most massive gravitationally bound structures in our Universe, forming at the intersections of cosmic filaments where matter flows converge \citep{Press1974}. As significant concentrations of matter, they serve as effective probes of both cosmological models and the physics of large-scale structure formation \citep{Kaiser1986}.

The high sensitivity of galaxy clusters to their formation history and initial conditions makes them excellent probes for evaluating our posterior resimulations. Agreement between observed and simulated clusters—in both spatial alignment and total mass—reinforces confidence that our posterior reconstructions faithfully represent the local Universe. This correspondence not only validates our methodology but also provides insights into the complex environmental conditions that governed the formation and evolution of these massive structures within the cosmic web.

Our analysis focuses on fourteen prominent galaxy clusters within the local supervolume: Abell 119, Abell 496, Abell 548, Coma, Hercules A2063, Hercules A2147, Hercules A2199, Leo, Norma, Perseus, Shapley A3571, Centaurus, Hydra, and Virgo. These systems span diverse environments—from relatively isolated haloes to dense supercluster regions like Shapley and Hercules—and were selected primarily for their overlap with previous constrained studies \citep{Stopyra2021,Pfeifer2023} and the availability of multiple independent observational mass estimates. This makes them ideal benchmarks for evaluating the accuracy of our reconstructions.

We examine two fundamental properties to assess reconstruction quality: (1) the spatial correlation between simulated and observed clusters, and (2) their total halo masses, quantified using \masscrit. Together, these metrics provide a robust test of how accurately \manticorelocal has inferred the distribution of matter in the local Universe.

\subsubsection{Identifying cluster counterparts}
\label{sect:identify_cluster_counterparts}

Establishing a robust correspondence between observed galaxy clusters and their simulated counterparts in constrained realizations presents a significant methodological challenge. The complex nature of cluster environments--where multiple massive haloes may exist in close proximity, observational uncertainties affect cluster positions, and varying mass estimation techniques introduce discrepancies--necessitates careful consideration of identification approaches.

Several methods have been employed in previous studies. The simplest approach, used by \citet{Stopyra2024_COLA} and \citet{2022MNRAS.512.5823M}, assigns each observed cluster to the most massive halo found within a fixed search radius (typically $\approx 30$~Mpc) of the observed position. While effective for isolated, massive clusters (${\sim}10^{15}$~\msol), this method becomes problematic in dense environments such as the Hydra-Centaurus and Shapley superclusters, where multiple candidate haloes can lead to ambiguous identifications.

The Simulating the Local Web (SLOW) simulations \citep{SLOW2024} address this limitation through a hierarchical selection process. Their method first matches observed clusters to groups in Tully's North/South Catalogue \citep{Tully2015} based on sky position and redshift, then searches for simulated counterparts within an expanding radius up to 65~Mpc. When multiple candidates remain, additional selection criteria—such as X-ray temperature, luminosity, and dynamical state—help determine the most likely counterpart. This approach improves reliability in complex environments but remains constrained by predetermined spatial thresholds.

An alternative framework developed by \citet{Pfeifer2023}, the Local Universe Model (LUM), offers a probabilistic solution to counterpart identification. Rather than imposing fixed search radii, LUM evaluates a probability function $p(r, M_{\mathrm{200,crit}})$ representing the likelihood of finding a halo with mass \masscrit at separation $r$ from a randomly chosen reference point in an unconstrained cosmological simulation\footnote{Note that the probability function \( p(r, M_{200,\mathrm{crit}}) \), as defined in \citet{Pfeifer2023}, assumes $r$ in units of \mpch and \masscrit in units of \msolh. We adopt the same unit convention when evaluating this function to ensure consistency with their definition.}. This formulation provides a null hypothesis test: if a massive halo appears near an observed cluster with low probability under random expectation, it becomes a strong counterpart candidate. The approach circumvents arbitrary spatial thresholds by selecting haloes with the lowest probability values as statistically significant matches.

Our study adopts the LUM framework for identifying cluster counterparts. The process involves three key steps: First, we convert the right ascension, declination, and redshift of each observed cluster into comoving Cartesian equatorial coordinates aligned with our posterior resimulation box. Second, we compute the probability value for every possible halo-cluster pair by evaluating the separation $r$ and mass $M_{\mathrm{200,crit}}$ for all simulated haloes above $1.47 \times 10^{14}$~\msol \citep[i.e., the $10^{14}$~\msolh cut from][]{Pfeifer2023}. This generates a probability matrix quantifying the likelihood of association between each observed and simulated cluster. Third, we apply the Hungarian algorithm \citep{Kuhn1955}, a combinatorial optimization method, to ensure a one-to-one correspondence that minimizes the total sum of probability values across all pairings.

While the LUM approach provides statistical rigour, its results remain sensitive to uncertainties in the assumed positions of observed clusters and to the chosen lower mass threshold for including simulated haloes. Redshift-space distortions and peculiar velocities affect the inferred distances of nearby clusters, particularly systems like Virgo where deviations from the Hubble flow can be significant \citep[see, e.g., Figure 3 in][]{SLOW2024}. To ensure consistency with previous studies, we adopt supergalactic coordinates from \citet{Pfeifer2023}—based on the Cosmicflows-3 Catalogue \citep{Tully2016}—for clusters included in their analysis. For the remaining clusters, we determine positions using right ascension, declination, and redshift values from the NASA/IPAC Extragalactic Database. Table \ref{tab:cluster_positions} presents the final assumed positions for all observed clusters in this study. We also apply the same lower mass threshold chosen by \citet{Pfeifer2023}—$10^{14}$~\msolh—for selecting simulated haloes.

\begin{table}
\centering
\begin{tabular}{|c|c|c|c|}
\hline
\textbf{Cluster} & \textbf{SGX} & \textbf{SGY} & \textbf{SGZ}\\
- & [Mpc] & [Mpc] & [Mpc] \\
\hline
Virgo Cluster & -4.8700 & 21.2979 & -0.896\\
Centaurus (A3526) & -50.6848 & 22.0594 & -11.1859\\
Hydra (A1060) & -35.8866 & 30.7912 & -36.362\\
Perseus (A426) & 72.5443 & -15.5785 & -18.8577\\
Coma (A1656) & 0.6561 & 104.4230 & 15.4200\\
Norma (A3627) & -72.9389 & -10.5106 &  8.9307\\
Leo (A1367) & -3.5077 & 98.9637 & -18.3534\\
\hline
Abell 119 & 81.8756 & -172.6919 & -2.1330\\
Abell 496 & 35.7520 & -73.7681 & -120.3062\\
Abell 548 & -8.5434 & -58.709 & -175.2079\\
Hercules (A2199) & 26.3204 & 82.7682 &104.2647\\
Hercules (A2147) & -37.1809 & 102.8894 &123.8503\\
Hercules (A2063) & -57.4741 & 109.7154  &96.2186\\
Shapley (A3571) & -154.9777  & 84.1743 &6.4640\\
\hline
\end{tabular}
\caption{Assumed three-dimensional supergalactic coordinates of the fourteen observed clusters analyzed in this study. The first seven clusters adopt values directly from Table 1 of \citet{Pfeifer2023}, ensuring consistency with their study. The remaining eight clusters have supergalactic positions derived using right ascension, declination, and redshift-distance measurements from the NASA/IPAC Extragalactic Database.}
\label{tab:cluster_positions}
\end{table}

\subsubsection{Detection Significance}
\label{sect:detection_significance}

\begin{figure}
    \includegraphics[width=\columnwidth]{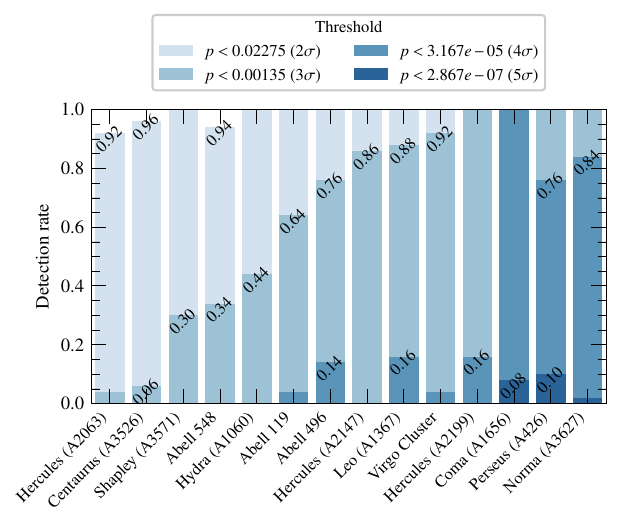}
    \caption{Detection rates of cluster counterparts across posterior resimulations, evaluated at four significance thresholds corresponding to $p$-values equivalent to $2\sigma$, $3\sigma$, $4\sigma$, and $5\sigma$ confidence. The clusters are ordered from left to right by increasing $3\sigma$ $p$-value—that is, from the most to the least confidently detected systems. These results highlight the statistical robustness of the reconstructed cluster population, ruling out random configurations and chance alignments.}
    \label{fig:detection_rate}
\end{figure}

Having identified cluster counterparts using the LUM framework, we now evaluate the statistical significance of these matches. The probability function $p(r, M_{\mathrm{200,crit}})$ not only guides our counterpart selection but also provides a direct measure of detection significance, allowing us to distinguish between meaningful recovery of observed structures and random alignments.

Following \citet{SLOW2024}, we define detection significance as $1 - p$, where $p$ represents the probability of finding a comparable halo in an unconstrained $\Lambda$CDM volume. Lower $p$-values correspond to higher detection significance. For instance, a $p$-value below 0.1587 indicates detection significance exceeding 80\% ($\approx 1\sigma$). Given the highly nonlinear relationship between $p$-values and confidence levels, we report the $p$-values directly rather than using approximate significance thresholds.

Our analysis of 700 identified counterparts across 50 posterior samples for 14 clusters reveals a median $p$-value of $5.0 \times 10^{-4}$, with a $10^{\text{th}}$-$90^{\text{th}}$ percentile range of $4.0 \times 10^{-6}$ to $6.6 \times 10^{-3}$. This corresponds to a median detection significance of $3.3\sigma$, with a $10^{\text{th}}$-$90^{\text{th}}$ percentile range of $2.5\sigma$ to $4.5\sigma$. These results demonstrate the high statistical significance of our reconstructed cluster population, effectively ruling out random configurations and chance alignments.

To assess the consistency of detection across our posterior distribution, we adopt the ``detection rate'' metric proposed by \citet{Pfeifer2023}, which measures the fraction of posterior resimulations in which a cluster counterpart exceeds a specific significance threshold. Figure \ref{fig:detection_rate} shows these detection rates at four significance levels, corresponding to confidence intervals of $2\sigma$, $3\sigma$, $4\sigma$, and $5\sigma$. All 700 cluster counterparts exceed $1\sigma$ significance across the full posterior distribution, and nearly all ($\gtrsim 99\%$) exceed $2\sigma$. The most massive systems--Coma, Perseus, and Norma--frequently achieve $4\sigma$ confidence in a substantial fraction of posterior realizations.

These results highlight the robust recovery of massive structures in our posterior resimulations. For comparison, the constrained simulation suite presented in \citet{Pfeifer2023} reported detection rates of $6-35\%$ at $\approx 2\sigma$ significance. Our substantially higher detection rates suggest that our inference process provides a more stable reconstruction of the local cluster population (see also the direct comparison in \cref{fig:detection_rate_comparison}).

While our detection significance analysis demonstrates high confidence in the reconstructed clusters, we acknowledge certain limitations. This analysis does not fully account for observational uncertainties in cluster mass or position when computing detection rates. Although we apply the same mass threshold of $M_{\mathrm{200,crit}} > 10^{14}$~\msolh as \citet{Pfeifer2023} to partially address mass-related uncertainties, we use fixed observed positions without incorporating positional error propagation. In reality, uncertainties in redshift-based distance and peculiar velocities affect cluster positions, and will influence computed $p$-values \citep[see, e.g., Figure 3 in][]{SLOW2024}. However, to maintain direct comparability with \citet{Pfeifer2023}, we follow their approach of using fixed observed positions.

Future work could enhance this analysis by incorporating observational uncertainties into the counterpart assignment process, providing more realistic estimates of detection significance. Additionally, rather than treating each cluster independently, a more comprehensive approach would assess the entire cluster population collectively through hierarchical Bayesian modelling or population-level likelihood analysis, offering a more robust statistical framework for evaluating the fidelity of posterior resimulations. As an intermediate step, it would be informative to analyze the correlation structure of detection significances across clusters—i.e., whether the recovery of one cluster systematically affects the recovery of others. Moreover, the current implementation of the LUM method treats clusters as point-like objects, neglecting their spatial extent. Extending the model to account for the finite size and morphology of massive haloes—e.g., through matching within virial radii or extended density structures—would improve the physical realism of the matching procedure and further refine significance estimates.

For subsequent analyses in this paper, we restrict our sample to cluster counterparts with a significance of at least $2\sigma$, which excludes fewer than 1\% of the total identified counterparts across our posterior resimulations.

\subsubsection{Cluster positions and velocities}
\label{sect:cluster_positions}

\begin{figure}
    \includegraphics[width=\columnwidth]{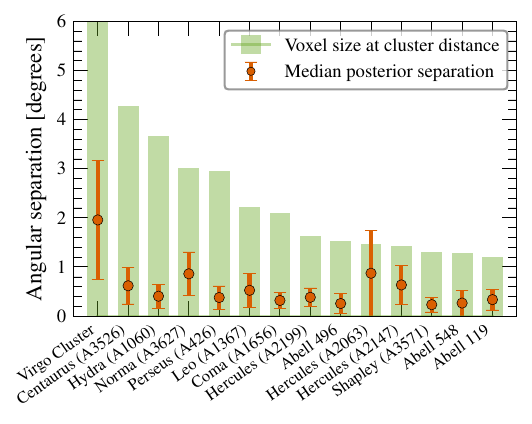}
    \caption{Median angular separation between the cluster counterparts and their observed positions across the posterior resimulations, with error bars indicating the \tentoninety percentile range. The shaded bars represent the angular size of a single voxel in the inference grid at the median cluster distance, providing a reference for the nominal resolution limit. Despite this constraint, the inferred cluster positions exhibit significantly improved alignment, demonstrating the inference process's ability to recover cluster locations with higher precision than the raw grid resolution alone would suggest.}
    \label{fig:cluster_angsep}
\end{figure}

\begin{figure*}
    \includegraphics[width=\textwidth]{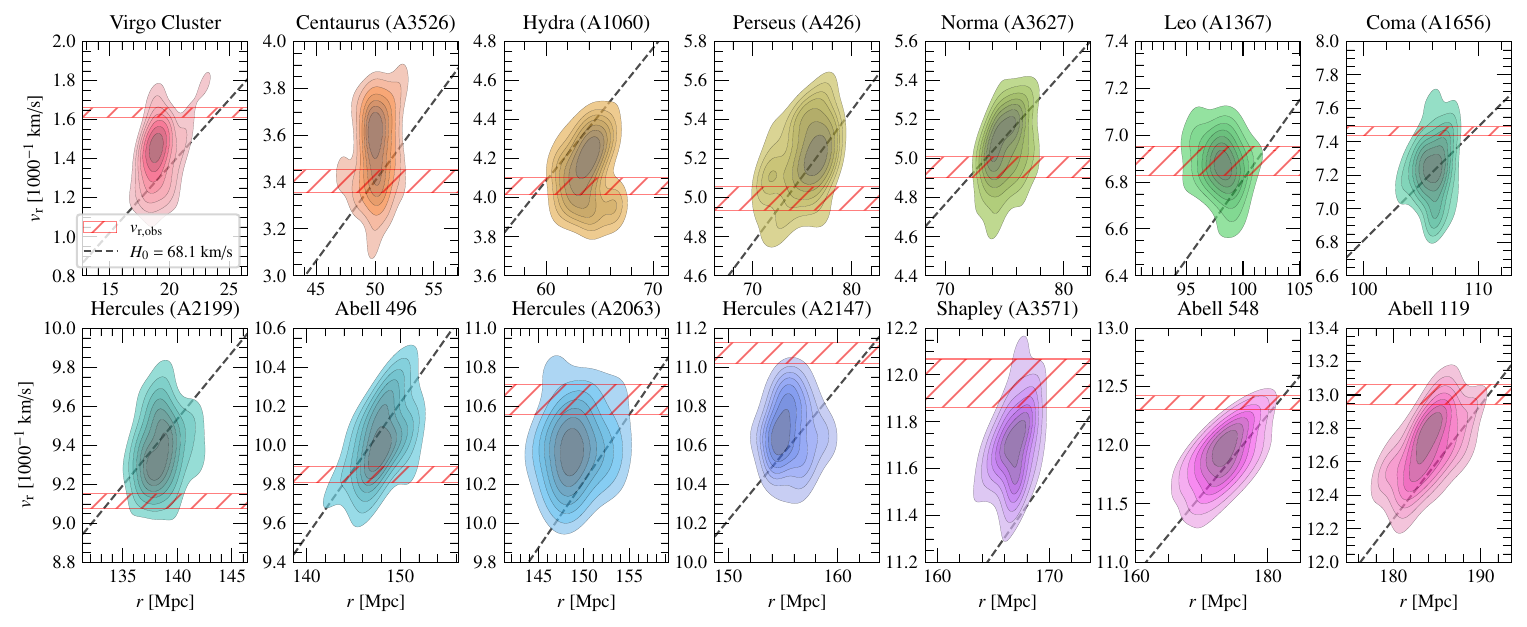}
    \caption{Two-dimensional posterior distributions of physical distance and recession velocity for the fourteen cluster counterparts. Each panel includes a line representing the Hubble relation from the assumed cosmology in the posterior resimulations. The shaded regions indicate the observed recession velocities from the NED Extragalactic Database. The strong agreement between the inferred and observed velocities demonstrates the ability of the constrained resimulations to recover the large-scale velocity field of the local Universe.}
    \label{fig:cluster_hubble}
\end{figure*}

The high detection significance of our cluster counterparts, as discussed in the previous section, depends on both their mass and spatial alignment with observed clusters. The impressive detection rates shown in \cref{fig:detection_rate} suggest strong positional correspondence, indicating that our inferred counterparts generally match well with observed locations.

From an observational perspective, galaxy clusters have exceptionally precise sky coordinates with negligible angular uncertainties. Consequently, any angular separation between inferred and observed positions primarily reflects limitations in our inference process rather than observational errors. Figure \ref{fig:cluster_angsep} presents the angular separation, in degrees, between our inferred cluster counterparts and their observed positions. The plot shows the median posterior separation for each cluster, with error bars indicating the 10th to 90th percentile range. For context, we also display the angular size of a single voxel in our inference grid (the angular extent subtended by $3.9$~Mpc at the median cluster distance), which presumably represents the fundamental resolution limit of our inference grid in the initial conditions.

A striking feature of these results is that median angular separations (typically 1--2 degrees) are significantly smaller than the voxel width (3--5 degrees). This indicates that cluster positions are reconstructed with precision beyond the nominal grid scale. This enhanced positional accuracy stems from the nature of Lagrangian regions in the initial conditions: proto-haloes that evolve into galaxy clusters are approximately an order of magnitude larger in the early Universe than their final virialized structures \citep[e.g.,][]{Bond1996,2024MNRAS.534.3120S,Doeser2024}. Consequently, large-scale structures associated with clusters span multiple voxels in the initial conditions, allowing their positions to be inferred with greater precision than the basic voxel resolution would suggest.

Additionally, observed clusters are typically traced by numerous galaxies distributed across multiple voxels in the inference grid. These galaxies, serving as constraints in our inference process, help refine the reconstructed cluster positions by leveraging information across broader regions. This effectively enhances the accuracy of inferred cluster locations beyond the raw voxel scale.

While angular alignment provides an important metric for evaluating reconstruction quality, an equally critical test is whether our inferred cluster counterparts correctly reproduce the observed redshift-distance relation \citep{Hubble1929}. Figure \ref{fig:cluster_hubble} presents the posterior distribution of recession velocity versus distance for each cluster, forming a Hubble diagram for direct comparison with observational expectations. The hatched horizontal bands indicate the observed cluster redshift and its associated uncertainty. A successful inference should place the posterior realizations within this range, demonstrating that reconstructed clusters not only align spatially but also exhibit dynamically consistent velocities.

The results show that nearly all the posterior distributions for the fourteen counterparts overlap with the expected velocity range, confirming the robustness of our resimulations in recovering the observed Hubble flow. Notably, this velocity reconstruction is achieved without direct velocity constraints in our inference process—the agreement arises purely from gravitational dynamics encoded in the forward model. This is particularly impressive for systems like Perseus, Coma, and Virgo, where inferred velocity distributions closely match observations despite their complex surrounding environments. The only exception is Hercules (A2147), which shows a posterior velocity distribution slightly below the observed range, suggesting the inferred cluster may be too close. However, this kind of discrepancy highlights the power of constrained realizations in identifying specific systems that warrant further investigation. The deviation in the cluster's recession velocity (or distance) could arise from enhanced dynamical interactions, unaccounted substructure, or systematic effects in our reconstruction process or the data. Understanding why this particular system differs while others match expectations provides an opportunity to probe the underlying assumptions of our inference framework and the local dynamics around this cluster.

Together, the results from both angular separation analysis and the Hubble diagram confirm the reliability of our constrained resimulations in reproducing the local cluster population. The strong spatial alignment demonstrates successful recovery of correct large-scale structure, while agreement in the Hubble diagram validates the inferred kinematics. The ability to recover both position and velocity information for each local cluster individually, without directly incorporating velocity constraints, underscores the predictive power of our approach and provides a robust foundation for investigating the formation and evolution of large-scale cosmic structures.

\subsubsection{Cluster masses}
\label{sect:cluster_masses}

\begin{figure}
    \includegraphics[width=\columnwidth]{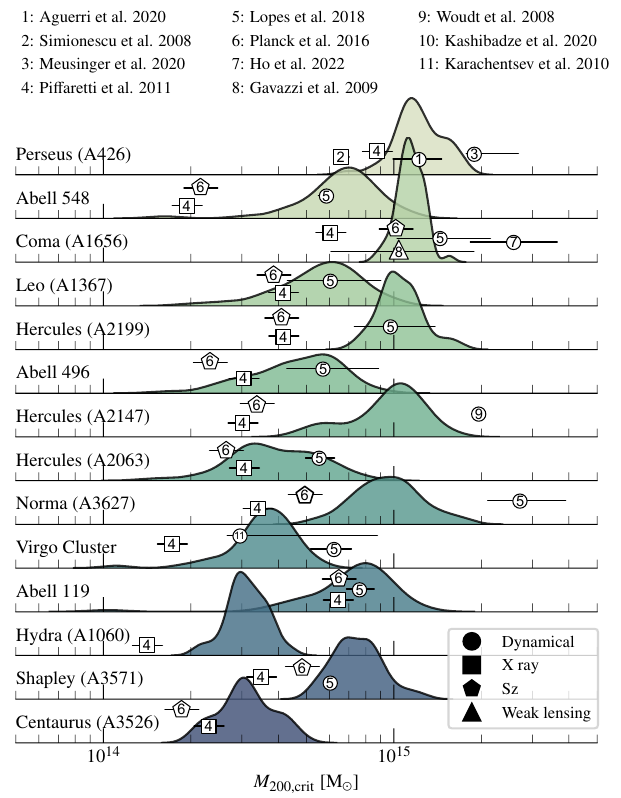}
    \caption{Posterior mass distributions for fourteen galaxy clusters in the local supervolume (shaded regions). Overlaid data points with error bars represent mass estimates from the literature, derived using various observational techniques, each distinguished by different marker shapes as indicated in the legend. The strong agreement between posterior distributions and observational estimates highlights the effectiveness of the \manticore model in accurately reproducing the masses of local clusters.}
    \label{fig:cluster_masses}
\end{figure}

\begin{figure*}
    \includegraphics[width=\textwidth]{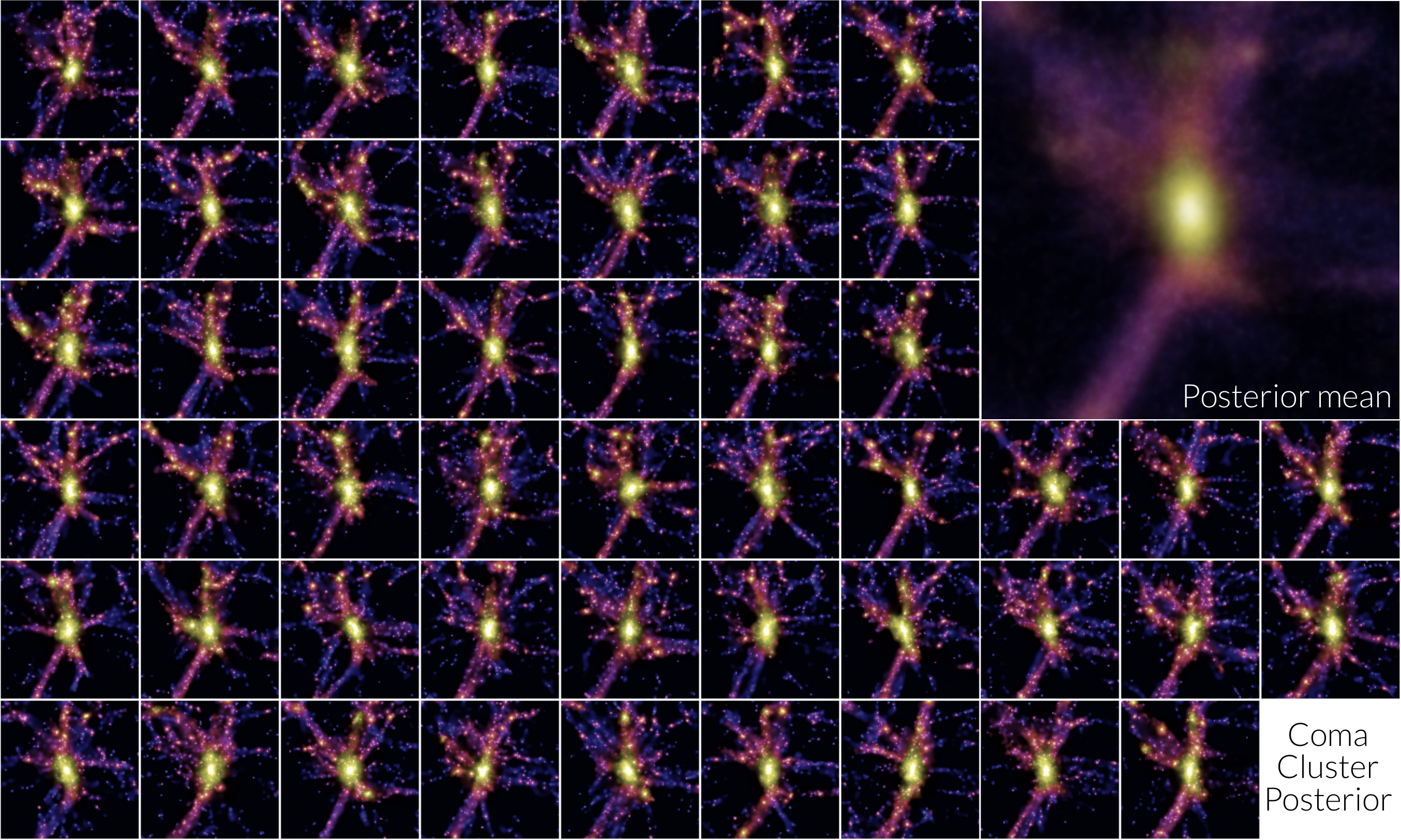}
    \caption{Posterior distribution of the Coma Cluster's dark matter structure, shown for all 50 samples. Each image is coloured by the dark matter density in log units. The posterior samples illustrate variations in fine-grain features around the cluster, yet the core and large-scale filamentary structures remain consistent across the samples. The inset image in the upper right shows the mean of the posterior samples, emphasizing the robust and coherent configuration of the core and filaments predicted to define the Coma Cluster. By visual inspection, there are predicted to be between 2 and 4 filaments connecting to the Coma cluster.}
    \label{fig:coma_posterior}
\end{figure*}

Astronomical estimation of galaxy cluster masses relies on four primary methodologies, each with distinct physical foundations and observational requirements. Dynamical methods \citep[e.g.,][]{Smith1936,Merritt1987} infer mass from galaxy motions within clusters, typically fitting observed velocity dispersions to a Navarro-Frenk-White \citep[NFW,][]{1997ApJ...490..493N} profile. These approaches assume dynamical equilibrium—an assumption that fails in unrelaxed or merging systems, potentially leading to mass overestimation \citep{2014MNRAS.441.1513O}.

Weak gravitational lensing \citep[e.g.,][]{Schneider1992} offers mass estimates independent of equilibrium assumptions by measuring distortions in background galaxy shapes caused by the cluster's gravitational field. While powerful, this method is susceptible to contamination from large-scale structure along the line of sight and requires careful discrimination between cluster members and unrelated galaxies \citep{2012NJPh...14e5018R}.

X-ray techniques \citep[e.g.,][]{Sarazin1986,Vikhlinin2006} utilize thermal bremsstrahlung emission from the intracluster medium, assuming hydrostatic equilibrium between hot gas and the cluster's gravitational potential. This assumption breaks down in merging clusters and regions influenced by AGN feedback, where non-thermal pressure support affects mass estimates. These biases can be partially mitigated by measuring gas properties out to large radii, such as $R_{200}$.

The Sunyaev-Zel'dovich (SZ) effect \citep{Sunyaev1970,Sunyaev1972,Sunyaev1980} measures spectral distortion of cosmic microwave background photons as they scatter off electrons in the intracluster gas. SZ-based mass estimates rely on scaling relations between the integrated Compton parameter ($Y_{\mathrm{SZ}}$) and total cluster mass, but uncertainties in the calibration of these relations and the hydrostatic mass bias introduce significant errors \citep{2007ApJ...668....1N}.

Our analysis builds on the cluster mass compilation by \citet{Stopyra2021}, which provides a detailed discussion of the advantages and limitations of various mass estimators. For the fourteen local clusters in this study, we compare mass estimates derived from multiple observational methods, presented in \cref{fig:cluster_masses}. Consistent with previous findings \citep[e.g.,][]{Wu1998, Lovisari2020, Stopyra2021}, substantial variations exist between different observational techniques, with dynamical estimates frequently yielding the highest mass values \citep[e.g.,][]{Foex2017}.

Field-level inference techniques provide an independent approach to constraining cluster masses. The posterior distributions for total halo masses predicted by the \manticore model are also shown in \cref{fig:cluster_masses}. Perseus emerges as the most massive system within $R \leq 200$ Mpc, with a median mass of $M = 1.18 \times 10^{15}$~\msol. Centaurus, by contrast, is the least massive cluster in our sample, with a median mass of $M = 3.04 \times 10^{14}$~\msol. 

The inferred mass posteriors exhibit sharp, well-defined peaks for most clusters, indicating well-constrained solutions with minimal degeneracies. Some clusters, such as Abell 496, show broader distributions spanning nearly an order of magnitude in mass. Nevertheless, all fourteen clusters remain within the range of observational mass estimates, demonstrating the effectiveness of the \manticore model in simultaneously reproducing the cluster population of the local supervolume through a self-consistent structure formation framework.

\subsection{Connectivity of the cosmic web around the Coma cluster}
\label{sect:coma_connectivity}

Galaxy clusters occupy the nodes of the cosmic web, residing at the intersections of large-scale filamentary structures \citep{Cautun2014}. The number and configuration of filaments connected to a cluster—its connectivity—plays a fundamental role in cluster growth and evolution by serving as conduits for continuous matter accretion. This connectivity provides valuable insights into a cluster's cosmic environment, influencing both its mass accretion history and the evolution of its galaxy population. Both theoretical and observational studies have established correlations between cluster connectivity and mass \citep{Codis2018, Sarron2019, Darragh2019}, with more massive haloes typically exhibiting higher connectivity. This relationship emerges naturally in hierarchical structure formation models, where larger systems grow through mergers of smaller structures and steady inflow of matter along connected filaments \citep{Gouin2021}.

The standard metric for quantifying connectivity is the filament number, $\kappa$, which counts the filaments intersecting a cluster's virial region \citep{Sarron2019}. Observational studies using filament identification methods such as DisPerSE \citep{Sousbie2011} or Minimum Spanning Trees \citep{Alpaslan2014} find that galaxy clusters typically exhibit connectivity values of 2 to 5, increasing with halo mass \citep{Malavasi2020, Arag2010}. Numerical simulations predict similar trends \citep{Codis2018, Gouin2021}, supporting the interpretation that connectivity traces both the dynamical state of a cluster and its role within the larger cosmic structure.

The Coma cluster—one of the most massive and thoroughly studied systems in the local Universe—provides an excellent test case for connectivity studies. Galaxy redshift surveys and density field reconstructions estimate Coma's connectivity at approximately 2.5, with at least three dominant filaments extending outward \citep{Malavasi2020, Mahajan2018}. These structures have been detected across multiple observational domains, including galaxy distributions, X-ray emissions, and Sunyaev-Zel'dovich measurements \citep{Gouin2021, Tanimura2019}, suggesting that they actively channel gas into the cluster and contribute to its ongoing mass assembly.

Our posterior resimulations offer a unique opportunity to examine Coma's connectivity from a field-level inference perspective. Figure \ref{fig:coma_posterior} presents the dark matter distribution in a 75 × 75 × 25 Mpc region surrounding the Coma cluster, along with the posterior mean of these realizations in the upper right corner. While individual realizations show unique fine-scale structures, the large-scale features—including the main filaments and the cluster core—remain consistent across samples. The posterior mean smooths out small-scale variations, preserving only the dominant features of the cosmic web. Visual inspection suggests the presence of two to four filaments feeding into Coma, aligning with observational estimates \citep{Malavasi2020, Sarron2019}.

A comprehensive quantitative analysis of filament connectivity in \manticorelocal will be presented in future work. However, the ability to directly compare reconstructed dark matter structures with observationally inferred filamentary networks demonstrates the power of these simulations for testing structure formation models. Extending these analyses to our broader sample of galaxy clusters will enable detailed statistical studies of connectivity and its implications for mass assembly within the cosmic web.

In summary, our \manticorelocal posterior resimulations demonstrate exceptional fidelity in reproducing the cluster population of the local Universe. The fourteen prominent clusters investigated are robustly recovered across the posterior, with high detection significance, precise positional alignment, dynamically consistent velocities, and total halo masses in agreement with multi-wavelength observational estimates. The success of the reconstruction is particularly notable given that no direct velocity or mass constraints were imposed—these properties emerge purely from the gravitational dynamics and observational data encoded in the field-level inference. This agreement across spatial, dynamical, and mass dimensions provides compelling evidence that our constrained realizations faithfully capture the large-scale structure and environmental context of the nearby Universe, validating the \manticore framework as a powerful tool for cosmological inference and forward modelling of local structure.

The posterior properties for the fourteen clusters of \manticorelocal are reported in \cref{tab:cluster_properties}.

\begin{table}
\centering
\begin{tabular}{@{}l@{\hspace{6pt}}c@{\hspace{6pt}}c@{\hspace{6pt}}c@{\hspace{6pt}}c@{}}
\hline
Cluster & \masscrit & $r$ & $v_r$ & $p$-value \\
 & [$10^{14} M_{\odot}$] & [Mpc] & [$10^3 \text{ km s}^{-1}$] & [$10^{-3}$] \\
\hline
Abell 119 & $7.72^{+1.77}_{-3.00}$ & $183.83^{+4.21}_{-3.75}$ & $12.69^{+0.26}_{-0.29}$ & $0.755^{+5.460}_{-0.627}$ \\[6pt]
Abell 496 & $4.74^{+1.75}_{-1.96}$ & $147.85^{+3.26}_{-2.66}$ & $10.02^{+0.25}_{-0.23}$ & $0.392^{+6.170}_{-0.364}$ \\[6pt]
Abell 548 & $6.91^{+1.83}_{-2.48}$ & $173.65^{+9.06}_{-2.45}$ & $11.94^{+0.46}_{-0.26}$ & $1.800^{+6.040}_{-1.200}$ \\[6pt]
Centaurus (A3526) & $3.04^{+1.18}_{-0.74}$ & $50.00^{+1.71}_{-1.29}$ & $3.56^{+0.17}_{-0.21}$ & $5.580^{+6.960}_{-4.010}$ \\[6pt]
Coma (A1656) & $11.36^{+1.45}_{-1.21}$ & $106.13^{+1.48}_{-1.64}$ & $7.24^{+0.22}_{-0.26}$ & $0.003^{+0.009}_{-0.002}$ \\[6pt]
Hercules (A2063) & $3.75^{+2.24}_{-1.02}$ & $148.79^{+4.48}_{-2.15}$ & $10.40^{+0.26}_{-0.24}$ & $6.330^{+5.970}_{-3.660}$ \\[6pt]
Hercules (A2147) & $10.02^{+2.64}_{-4.20}$ & $155.43^{+2.56}_{-1.56}$ & $10.65^{+0.20}_{-0.17}$ & $0.520^{+1.730}_{-0.369}$ \\[6pt]
Hercules (A2199) & $10.27^{+2.45}_{-1.71}$ & $138.27^{+2.00}_{-1.58}$ & $9.36^{+0.25}_{-0.18}$ & $0.116^{+0.284}_{-0.099}$ \\[6pt]
Hydra (A1060) & $3.14^{+0.69}_{-0.58}$ & $63.95^{+2.34}_{-2.70}$ & $4.15^{+0.20}_{-0.21}$ & $1.810^{+4.920}_{-1.610}$ \\[6pt]
Leo (A1367) & $5.84^{+1.81}_{-2.18}$ & $98.28^{+1.34}_{-2.32}$ & $6.87^{+0.14}_{-0.12}$ & $0.120^{+1.350}_{-0.096}$ \\[6pt]
Norma (A3627) & $9.55^{+3.81}_{-2.52}$ & $74.95^{+1.96}_{-1.40}$ & $5.05^{+0.20}_{-0.26}$ & $0.011^{+0.030}_{-0.008}$ \\[6pt]
Perseus (A426) & $11.82^{+3.77}_{-1.74}$ & $75.47^{+1.98}_{-3.70}$ & $5.18^{+0.24}_{-0.26}$ & $0.003^{+0.098}_{-0.003}$ \\[6pt]
Shapley (A3571) & $7.39^{+2.00}_{-1.57}$ & $166.89^{+1.47}_{-2.18}$ & $11.68^{+0.23}_{-0.18}$ & $1.830^{+2.480}_{-1.020}$ \\[6pt]
Virgo Cluster & $3.59^{+1.00}_{-0.90}$ & $19.03^{+1.87}_{-0.89}$ & $1.44^{+0.23}_{-0.22}$ & $0.437^{+0.507}_{-0.315}$ \\
\hline
\end{tabular}
\caption{Properties of galaxy clusters from \manticorelocal, showing median values with asymmetric error bars where the superscript represents the difference between the 90th percentile and median, and the subscript represents the difference between the median and 10th percentile.}
\label{tab:cluster_properties}
\end{table}
\section{Discussion}
\label{sect:discussion}

\subsection{Comparison to previous work}
\label{sect:comparison_to_previous_work}

\begin{figure}
    \includegraphics[width=\columnwidth]{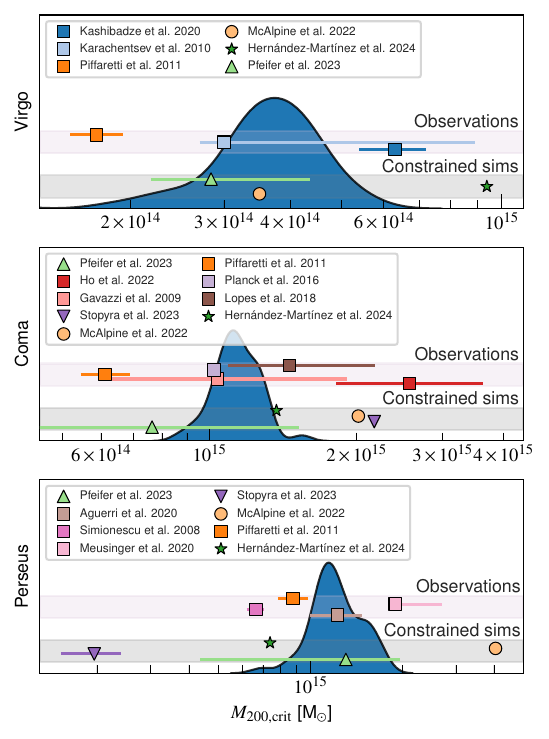}
    \caption{Posterior mass distributions for the Virgo, Coma, and Perseus clusters from \manticore (blue distributions), compared with observational mass estimates (coloured markers enclosed in purple-shaded regions) and predictions from previous constrained simulations (coloured markers enclosed in grey-shaded regions). \manticore achieves agreement with observations across all three clusters—an improvement over earlier efforts that struggled to reproduce the full set consistently. The narrowness of the posterior distributions (coefficients of variation $\approx 0.2$-0.3) begins to meaningfully discriminate between competing observational estimates, highlighting potential systematic biases in some methods, particularly X-ray-based measurements.}
    \label{fig:mass_comparison}
\end{figure}

\begin{figure}
    \includegraphics[width=\columnwidth]{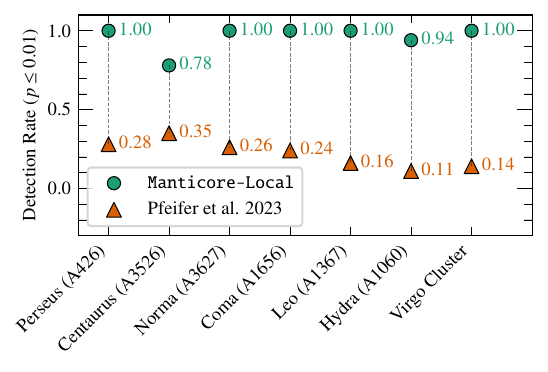}
    \caption{The detection rates (see \cref{sect:detection_significance}) for seven local clusters common to this work and that of \citet{Pfeifer2023}. Only the lower-mass Centaurus (78\%) and Hydra (94\%) systems do not have 100\% detection rates across the \manticorelocal posterior. In contrast, \citet{Pfeifer2023} report significantly lower detection rates, ranging from 11\% to 35\%, underscoring the enhanced robustness and reproducibility of the \manticore methodology.}
    \label{fig:detection_rate_comparison}
\end{figure}

Constrained simulations have historically struggled to simultaneously reproduce the full local cluster population with both accuracy and consistency. These challenges have spanned methodologies, from reconstructions based on galaxy distributions to those using peculiar velocity fields.

Within reconstructions based on galaxy distributions, \citet{2022MNRAS.512.5823M} reported systematically inflated cluster masses—sometimes by factors of 2–3—relative to observational constraints. The abundance of clusters with \masscrit $\geq 10^{15}$~\msolh in their reconstruction was substantial enough to place the \lcdm model in tension with local cluster counts according to the metric defined by \citet{Stopyra2021}. This discrepancy was later attributed to inaccuracies in the gravitational modelling approach employed by \citet{Jasche2019}, which introduced systematic biases. Specifically, \citet{Stopyra2024_COLA} demonstrated that the approximate gravity solver used in that earlier inference distorted the inferred Lagrangian protohalo regions, inflating cluster masses when subsequently evolved with full $N$-body dynamics. By adopting the more precise \texttt{COLA} solver, \citet{Stopyra2024_COLA} significantly mitigated these biases. Nevertheless, residual discrepancies persisted, notably with the masses of the Perseus and Norma clusters still underestimated by approximately 20–80\%.

Reconstructions utilizing peculiar velocity constraints have also advanced significantly, although notable challenges persist. \citet{Pfeifer2023}, through their Local Universe Model (LUM), developed an ensemble of constrained realizations that successfully captures many large-scale aspects of the observed cosmic velocity field and reliably identifies prominent galaxy clusters. However, their predictions for individual cluster masses exhibit considerable uncertainty, often exceeding 50\% of the mean, limiting their ability to clearly distinguish between competing observational estimates or identify measurement systematics.

These uncertainties likely arise from several compounding factors. The Cosmicflows peculiar velocity dataset, which underpins these simulations, is inherently sparse, noisy, and affected by observational biases, compromising the accuracy of small-scale structure reconstruction. Furthermore, the CLUES pipeline—employing a Wiener-filtered late-time velocity field and a reverse Zeldovich approximation followed by constrained realizations—does not fully sample the Bayesian posterior distribution of initial conditions given the data. As detailed by \citet{2013MNRAS.430..888D,2013MNRAS.430..902D,2013MNRAS.430..912D}, such methods effectively recover large-scale modes but struggle to propagate non-linear uncertainties or fully capture mode correlations at smaller scales. Consequently, while the ensemble provides useful insights into large-scale structure, it does not represent a statistically calibrated posterior distribution in the Bayesian sense, limiting its interpretability and the robustness of derived uncertainty estimates.

Hydrodynamic simulations in the field are starting to make progress. The \texttt{SLOW} simulation \citep{SLOW2024}, a hydrodynamical resimulation based on the \texttt{CLONES} initial conditions \citep{Sorce2018}, has generally achieved good agreement with observed cluster masses, though discrepancies remain. In particular, the Virgo cluster's mass is significantly overestimated, by a factor of 2–5. This overestimation likely contributes to the pronounced overdensity near Virgo and the pronounced underdensity in its foreground reported by \citet{Dolag2023}, contrasting with the smoother radial density profile inferred by \manticorelocal (see \cref{fig:radial_density_profile}). We note that the methodology for initial conditions generation behind the \texttt{CLONES}/\texttt{SLOW} simulations and the simulations presented by \citet{Pfeifer2023} are effectively identical.

In \cref{fig:cluster_masses}, the \manticorelocal resimulations demonstrated excellent consistency with observed cluster masses in the local Universe. This comparison is further refined in \cref{fig:mass_comparison}, which highlights mass predictions for three prominent clusters—Virgo, Coma, and Perseus—alongside earlier constrained studies. In addition to their alignment with observational estimates, the \manticorelocal results exhibit notably narrower posterior distributions, quantified by the coefficient of variation\footnote{Defined as $\mathrm{CV} = (P_{90} - P_{10}) / (2 \times \mathrm{median})$, where $P_{10}$ and $P_{90}$ denote the 10th and 90th percentiles of the posterior, respectively.}, with $\mathrm{CV} = 0.35$, $0.21$, and $0.19$ respectively. This precision stands in marked contrast to the broader mass predictions from \citet{Pfeifer2023} ($\mathrm{CV} = 0.38$, $0.71$, and $0.58$) and significant outliers found by both \citet{Stopyra2024_COLA} and the \texttt{SLOW} simulation.

A complementary assessment involves the \textit{detection rate}, defined as the fraction of posterior realizations in which a given cluster is detected with significance surpassing a specified threshold (see also \cref{sect:detection_significance}). \cref{fig:detection_rate_comparison} compares detection rates at a significance level corresponding to $p \leq 0.01$ (just under $2\sigma$) for seven clusters analyzed by both \manticorelocal{} and \citet{Pfeifer2023}. \manticorelocal achieves a 100\% detection rate for five of these clusters, with only the lower-mass Centaurus and Hydra systems showing slightly reduced, though still high, rates of 78\% and 94\%, respectively. Conversely, \citet{Pfeifer2023} reported notably lower detection rates, ranging from just 11\% to 35\%, underscoring the greater robustness and reproducibility of the \manticore{} methodology.

This enhanced precision enables critical examination of variations among observational mass estimates. For instance, several X-ray-derived cluster masses shown in \cref{fig:cluster_masses} systematically fall below the posterior medians inferred by \manticorelocal. This discrepancy potentially reflects hydrostatic mass biases or non-equilibrium gas dynamics, both known sources of systematic error in X-ray mass measurements. Our precise field-level posterior estimates—particularly when extended to future hydrodynamical resimulations—can thus serve as valuable benchmarks for isolating these observational biases and guiding targeted follow-up studies.

The progression from \citet{2022MNRAS.512.5823M} to \citet{Stopyra2024_COLA} to \manticorelocal—using the same underlying data and \borg inference framework—demonstrates how methodological refinements, particularly in gravity modelling, noise calibration, and posterior sampling, can dramatically improve reconstruction fidelity. \manticore incorporates a physically consistent structure formation model, a flexible galaxy bias prescription, and rigorous field-level inference to extract statistically robust constraints. Compared to peculiar velocity-based efforts like \citet{Pfeifer2023} and hydrodynamic resimulations like \texttt{SLOW}, our approach delivers tighter, more consistent predictions for the local cluster population.

\subsection{Coma cluster distance as a probe of the Hubble tension}
\label{sect:coma_distance_tension}

The distance to the Coma cluster offers a unique lens through which to examine the Hubble tension—the persistent discrepancy between local measurements of the Hubble constant ($H_0$) and values inferred from early-Universe probes such as the cosmic microwave background \citep{2024ARA&A..62..287V}. As highlighted by \citet{Scolnic2025}, Coma is among the most thoroughly studied galaxy clusters, with decades of high-quality distance measurements. It plays a central role in anchoring distance ladder calibrations and is now a key reference point for new cosmological probes, such as the fundamental plane relation from the \textit{Dark Energy Spectroscopic Instrument} (DESI) peculiar velocity survey \citep{Said2024}.

Coma’s distance has been estimated using a range of independent techniques—including Type Ia supernovae \citep{Scolnic2025}, surface brightness fluctuations \citep{Thomsen1997,Jensen1999}, and fundamental plane relations \citep{Hjorth1997}—yielding consistent results around $\approx 80{-}100$~Mpc. The convergence of these methods enables robust cross-validation, reducing the likelihood of major systematic errors in any single technique. At a redshift of $z \approx 0.023$, Coma is far enough to lie within the Hubble flow, yet close enough to permit accurate distance measurements—placing it in a sweet spot for cosmological inference.

The latest results from \citet{Scolnic2025} have reaffirmed the significance of the tension. Distance calibrations tied to the local distance ladder place Coma at $\approx 98$~Mpc, while inverse distance ladder methods—anchored in \textit{Planck}+\lcdm cosmology—place it at $\approx 112$~Mpc. This discrepancy corresponds to a $4.6\sigma$ deviation, demonstrating that the Hubble tension is not limited to a mismatch in $H_0$ alone but extends to the fundamental calibration of the cosmic distance scale. These distances differ by more than 10\%, far exceeding the quoted measurement errors, and highlight a critical need for independent cross-checks.

Field-level constrained simulations offer such an independent test. Unlike traditional approaches that apply external peculiar velocity corrections to Coma’s redshift, our method reconstructs the full dynamical context of the local Universe from initial conditions. The \manticorelocal posterior resimulations reproduce the actual large-scale structure—including Coma and its surrounding cosmic web—while maintaining global consistency with a \lcdm cosmology. These reconstructions provide a forward-modeled, dynamical estimate of the Coma cluster’s position and velocity—one that does not rely on redshift-based corrections or velocity priors. Instead, Coma’s motion arises self-consistently from the gravitational evolution of the reconstructed local density field, constrained by galaxy observations.

\begin{figure}
    \includegraphics[width=\columnwidth]{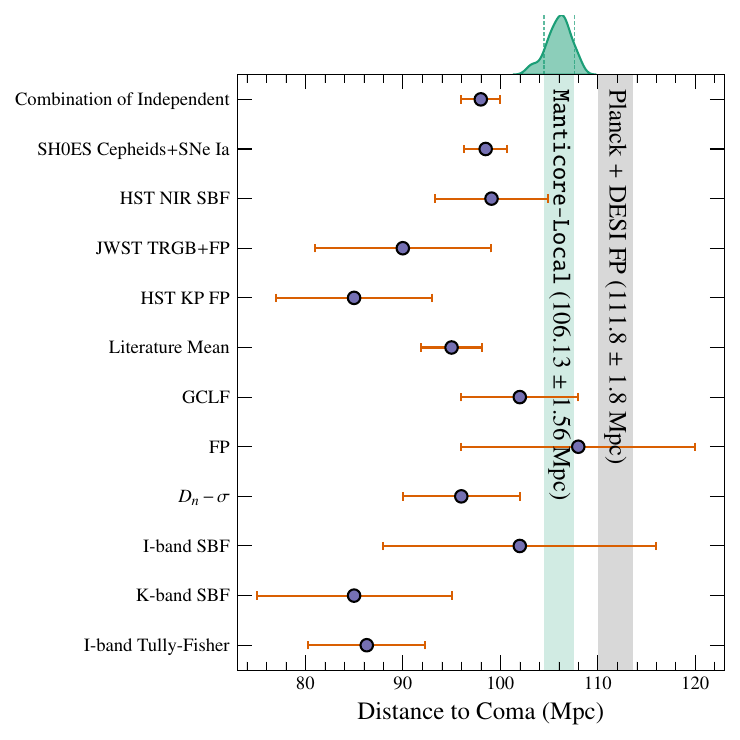}
    \caption{Distance estimates to the Coma cluster from various observational tracers (points with error bars), taken from Table 2 of \citet{Scolnic2025}. The green-shaded region indicates the posterior estimate from \manticorelocal\ (with the band showing the \tentoninety percentile range). The gray-shaded region corresponds to the \textit{Planck}-calibrated distance using the DESI fundamental plane relation. The \manticorelocal posterior lies between the two regimes, offering an independent dynamical prediction that may help disentangle the origin of the Hubble tension.}
    \label{fig:distance_to_coma}
\end{figure}

\Cref{fig:distance_to_coma} compares distance estimates from various observational tracers compiled in \citet{Scolnic2025} with the \manticorelocal posterior prediction. Our simulations yield a median Coma distance of approximately 106 Mpc—intermediate between the local ladder estimate ($\approx 98$ Mpc) and the \textit{Planck}+\lcdm expectation ($\approx 112$~Mpc). This non-trivial result suggests that the inferred dynamics of the local structure—constrained by gravitational evolution and large-scale density—naturally produce a Coma-like object at a distance between the two conflicting observational regimes.

It is important to emphasize, however, that our inference and posterior resimulations are conditioned on a fixed cosmological model, including an assumed value of the Hubble constant. This means that the predicted distances are not fully independent of $H_0$: under the \lcdm cosmology assumed, the Hubble flow is fixed, and so the predicted Coma distance is constrained to lie near its redshift-defined value. The spread in our posterior prediction reflects the range of peculiar velocities that naturally emerge for Coma across realizations—given its mass, environment, and position within the reconstructed cosmic web. As such, our result tells us whether a self-consistent realization of the local Universe—one that correctly matches Coma's mass (see \cref{fig:cluster_masses}), spatial location, and redshift (see \cref{fig:cluster_hubble})—predicts a significant departure from the Hubble flow at fixed $H_0$.

Our finding that Coma is not significantly offset from the expected Hubble-flow distance under this cosmology implies that, within the \lcdm framework, large peculiar velocities or environmental effects are insufficient to reconcile the observed low distance from direct indicators. Reconciling the tension would therefore require either a different value of $H_0$, or a departure from standard cosmological assumptions—such as modifications to the metric expansion or gravitational dynamics.

This makes constrained simulations a powerful and complementary probe of the Hubble tension. By jointly enforcing global consistency with \lcdm and local agreement with observed structure, they enable distance predictions that are free from traditional distance indicators yet dynamically grounded. Extending this approach to other clusters and applying forward modelling to observable proxies—such as supernova light curves or galaxy kinematics—will provide a richer understanding of the interplay between velocity fields, environmental effects, and cosmological distance scales.

\subsection{Toward standardized validation for constrained realizations}
\label{sect:setting_a_benchmark}

As constrained cosmological simulations become increasingly sophisticated, establishing a core set of shared validation metrics will significantly improve comparability and foster community-wide progress. Currently, the diversity of goals, methods, and evaluation criteria makes it difficult to objectively assess the relative strengths or limitations of different models. While heterogeneity is a strength, some baseline consistency is essential to ensure meaningful scientific comparisons.

We propose a tiered framework of standardized validation metrics to guide the evaluation of constrained realizations, where Tier 1 and 2 should be demonstrated whenever possible and metrics at the level of Tier 3 are our eventual goal:

\begin{itemize}
    \item \textit{Tier 1: Statistical consistency with cosmological models.} At a minimum, constrained realizations should demonstrate that the parent volume hosting the constrained region is statistically compatible with a standard cosmology such as \(\Lambda\)CDM. This includes validating initial condition Gaussianity, halo mass functions, power spectra, and bispectra. Crucially, uncertainty quantification should be rigorously incorporated to propagate observational noise through to derived cosmological observables.

    \textit{Tier 2: Fidelity to observed local structure.} Reproducing known local structures provides a stringent test of inference quality. Metrics such as the detection rate of galaxy clusters—quantified probabilistically via methods like the Local Universe Model \citep{Pfeifer2023}—enable fair comparisons across simulations. However, such comparisons require a community-standardized catalog of cluster positions and properties to avoid ambiguities in what is being measured. Ideally, this effort would move beyond one-point comparisons to statistical evaluations of the entire cluster population.

    A complementary metric is the Bayesian evidence for the reconstructed peculiar velocity field, as introduced in the ``Velocity Field Olympics'' by \citet{Stiskalek2025}. This framework offers a robust, likelihood-based comparison across methods using direct observational datasets (e.g., Tully-Fisher and supernovae-based velocities), and has already demonstrated its value in distinguishing performance across successive generations of constrained simulations.

    \item \textit{Tier 3: Physically enriched observational comparisons.} As hydrodynamical simulations become increasingly feasible, additional comparisons become possible. These include direct tests against observed Sunyaev-Zel’dovich and X-ray cluster profiles, stellar population properties, or supercluster morphology (e.g., \citealt{SLOW2024}). Other advanced benchmarks could include consistency with observed local voids, correlations with the Integrated Sachs-Wolfe effect, or detailed comparisons between inferred and observed galaxy distributions, including bias modelling and selection effects.
\end{itemize}

Beyond metrics that compare constrained predictions to observations, there are also emerging frameworks that assess the \textit{internal stability} of constrained realizations themselves. A recent example is the halo overlap metric proposed by \citet{2024MNRAS.534.3120S}, which quantifies the consistency of individual halo properties across a suite of resimulations sharing large-scale constraints. This method evaluates how robustly halos are reconstructed by tracking the overlap of their initial Lagrangian patches across realizations, offering a direct measure of how strongly the constraints pin down specific structures. Importantly, this approach distinguishes between the \textit{precision} of a reconstruction---i.e., how tightly a halo’s properties are predicted given the data---and its \textit{accuracy} relative to the actual Universe. Such internal consistency metrics provide a complementary axis of validation, helping to disentangle true predictive power from stochastic variability and thus advancing our ability to benchmark constrained cosmological models.

To make these comparisons practical, it will be crucial to standardize reference dataset and to openly communicate methodological trade-offs (e.g., resolution vs. coverage). Publishing benchmark results in parallel with simulation releases would help the field move toward transparency and reproducibility.

Upcoming and ongoing surveys (e.g., \textit{Euclid}, LSST, SPHEREx, SKA) will offer new opportunities and constraints. A community-driven effort to create shared validation protocols—perhaps in the form of a public repository or working group—would provide lasting value. While no single metric can capture all aspects of a model’s performance, establishing a set of commonly adopted standards will allow us to understand not just where our models succeed, but also where they fail—and thus help us progress as a community.

\section{Conclusion}
\label{sect:conclusions}

In this work, we presented \manticorelocal, the most advanced Bayesian digital twin of our cosmic neighbourhood constructed to date. Leveraging field-level Bayesian inference with data from the \tmpp galaxy survey, our model incorporates refined galaxy bias modelling, high-fidelity reconstructions, and rigorous posterior predictive validation. As a result, \manticorelocal accurately reproduces the spatial distribution, mass hierarchy, and velocity fields of observed cosmic structures within the local Universe, outperforming previous state-of-the-art constrained simulations and velocity field reconstructions across multiple key metrics. 

Critically, \manticorelocal enables statistically rigorous uncertainty quantification, with five independent Markov chains passing stringent Gelman-Rubin convergence tests, demonstrating, for the first time, fully converged posterior sampling across the entire reconstructed volume.
Taken together, these results demonstrate the power of physically informed inference frameworks to produce observationally anchored, cosmologically consistent reconstructions of structure formation in our local Universe.

The key outcomes of this study include:

\begin{itemize}
    \item \manticorelocal demonstrates comprehensive statistical consistency with \lcdm cosmological expectations, validated by tests of Gaussianity of initial conditions (\cref{fig:wnf_ps,fig:1d_pdf,fig:wnf_qq}), halo mass function comparisons (\cref{fig:parent_halo_mass}), and power spectrum and bispectrum analyses (\cref{fig:parent_power_spectrum}).

    \item Our reconstructions achieve strong spatial coherence with observed structures (\cref{fig:2mpp_with_dm}), capturing key features of the local environment such as the Virgo, Coma, and Perseus clusters. Radial density profiles confirm a clear transition from significant local features to large-scale homogeneity, consistent with \lcdm expectations (\cref{fig:radial_density_profile}).

    \item The normalization of the halo mass function in the local Universe is found to fluctuate at approximately ±5\% relative to the mean \lcdm prediction across the region $100 < R < 200$~Mpc (\cref{fig:inner_halo_mass_function}). There is predicted to be zero or one massive (\masscrit $\geq 10^{15}$~\msolh) cluster within $R < 200$~Mpc (\cref{fig:N_above_15}), in line with \lcdm expectations. Overall, the local supervolume  ($R<200$~Mpc) is not predicted to be statistically unusual in the context of \lcdm.

    \item \manticorelocal achieves the highest Bayesian evidence across all major peculiar velocity datasets—2MTF, SFI++, CF4, LOSS, and Foundation—demonstrating a preferred reconstruction of the local velocity field compared to previous state-of-the-art models (\cref{fig:velocity_evidence_compare}).

    \item Detailed analyses of fourteen major galaxy clusters demonstrate highly accurate recovery of their masses, positions, and velocities with robust statistical significance (\cref{fig:detection_rate,fig:cluster_angsep,fig:cluster_masses,fig:cluster_hubble}). This underscores the model’s capability for simultaneous and precise reconstruction of multiple gravitationally bound structures, overcoming long-standing challenges in constrained simulations to match both dynamical and structural properties of the observed Universe (\cref{fig:mass_comparison}).
\end{itemize}

The reconstructions presented here will complement upcoming data from next-generation observational surveys, including ESA’s \textit{Euclid} satellite mission \citep{2024arXiv240513491E}, the \textit{Vera C. Rubin Observatory’s Legacy Survey of Space and Time} \citep{2012arXiv1211.0310L}, the SPHEREX \textit{All-Sky Spectral Survey} \citep{2014arXiv1412.4872D}, and the \textit{Square Kilometre Array} \citep{2004NewAR..48..979C}. The synergy between our detailed local reconstructions and these future datasets will enrich cosmological analyses, enhance the precision of observational interpretations, and drive novel discoveries in both astrophysics and fundamental physics.

Future efforts will extend this work through the upcoming \manticoredeep project, applying advanced inference methodologies to the combined SDSS and BOSS galaxy surveys. Covering a constrained volume approximately 100 times larger than \manticorelocal within a $L=6000$~Mpc parent volume at $5.9$~\mpch resolution, \manticoredeep will bridge detailed local reconstructions with extensive cosmological surveys, significantly advancing our understanding of cosmic evolution and the nature of structure formation across cosmic time.

The initial conditions, posterior resimulations and reduced products from \manticorelocal will be publicly available after the publication of this work via the \href{https://www.cosmictwin.org/}{Manticore Project website}\footnote{\href{https://www.cosmictwin.org/}{https://www.cosmictwin.org/}}. These data offer significant potential for further exploration, such as resimulations incorporating hydrodynamics, zoom-in studies of specific structures like the Local Group or Coma cluster, and detailed investigations into galaxy formation and evolution. We warmly encourage collaborations and invite the cosmological community to leverage these resources to foster novel insights and advance our collective understanding of cosmic structure and dynamics. We believe \manticorelocal will serve as one of the foundational tools for precision cosmology in the local Universe.

\section*{Acknowledgements}

We thank Harry Desmond, Till Sawala, Ludvig Doeser and Adam Andrews for their useful inputs and discussions.

We acknowledge computational resources provided by the Swedish National Infrastructure for Computing (SNIC) at the PDC Center for High Performance Computing, KTH Royal Institute of Technology, partially funded by the Swedish Research Council through grant agreement no. 2018-05973. This work used the DiRAC@Durham facility managed by the Institute for Computational Cosmology on behalf of the STFC DiRAC HPC Facility (www.dirac.ac.uk). The equipment was funded by BEIS capital funding via STFC capital grants ST/K00042X/1, ST/P002293/1, ST/R002371/1 and ST/S002502/1, Durham University and STFC operations grant ST/R000832/1. DiRAC is part of the National e-Infrastructure. In addition, this work has made use of the Infinity Cluster hosted by Institut d'Astrophysique de Paris, and was granted access to the HPC resources of TGCC (Très Grand Centre de Calcul), Irene-Joliot-Curie supercomputer, under the allocations A0170415682 and SS010415380.

JJ and GL acknowledge support from the Simons Foundation through the Simons Collaboration on "Learning the Universe". This work was made possible by the research project grant "Understanding the Dynamic Universe," funded by the Knut and Alice Wallenberg Foundation (Dnr KAW 2018.0067). Additionally, JJ acknowledges financial support from the Swedish Research Council (VR) through the project "Deciphering the Dynamics of Cosmic Structure" (2020-05143) and GL acknowledges support from the CNRS IAE programme \dquotes{Manticore}. RS acknowledges financial support from STFC Grant No. ST/X508664/1, the Snell Exhibition of Balliol College, Oxford, and the CCA Pre-doctoral Program.

This research has made use of the NASA/IPAC Extragalactic Database (NED), which is funded by the National Aeronautics and Space Administration and operated by the California Institute of Technology.

This work is done within the Aquila Consortium\footnote{\url{https://www.aquila-consortium.org/}}, Virgo Consortium\footnote{\url{https://virgo.dur.ac.uk/}} and Learning the Universe Collaboration\footnote{\url{https://learning-the-universe.org/}}.

\section*{Data Availability}

Data products from the \manticorelocal and \manticoremini resimulations will be made publicly available following publication via the \href{https://www.cosmictwin.org/}{Manticore Project website}.
The initial release will include data products based on the $1024^3$ resimulations presented in this work and include the $z=0$ halo catalogues, gridded matter density fields, HEALPix sky maps, radial density profiles, and three-dimensional velocity fields.
In addition, the release will include resimulations with an effective resolution of $2048^3$ over the constrained \tmpp region and high cadence merger trees. The corresponding initial conditions will be released at a later stage. These datasets are intended to support a broad range of scientific applications, including hydrodynamical resimulations, structure formation studies, and targeted analyses of the local Universe.



\bibliographystyle{mnras}
\bibliography{example} 


\appendix

\section{The \borg algorithm}
\label{sect:borg_algorithm}

This section provides an overview of the \textit{Bayesian Origin Reconstruction from Galaxies}, \borg, algorithm as used in this study, including both standard components and new extensions introduced in the \manticore model. While not an exhaustive treatment, we highlight the core principles and innovations relevant for our field-level inference framework. For more comprehensive details on the original BORG algorithm, we refer the reader to \citet{Jasche2019}.

At its core, \borg\ solves the inverse problem of cosmic structure formation: given the present-day galaxy distribution, it seeks to infer the set of most plausible initial conditions that, under gravitational evolution, gave rise to the observed structures. This is achieved through a hierarchical Bayesian framework, in which the likelihood function relates the observed galaxy distribution to the underlying dark matter density field, while the prior provides statistical constraints on the primordial fluctuations. The posterior distribution over initial conditions is explored using advanced Markov Chain Monte Carlo sampling methods, ensuring that the full range of uncertainties, including those stemming from observational limitations and theoretical modelling, are properly propagated.

The key to \borg's forward modelling approach is its gravitational structure formation model, which maps the initial density field to the present-day evolved density field. This evolution is modelled using a choice of differentiable approximate gravity solver, such as Second-Order Lagrangian Perturbation Theory \citep[2LPT,][]{Bouchet1995}, Particle-Mesh $N$-body solvers \citep[PM,][]{Hockney1988}, $\texttt{COLA}$ based methods \citep{Tassev2013,Izard2016} and most recently a neural network-based field-level emulator \citep{2023ApJ...952..145J,Doeser2024}. While 2LPT provides an efficient and reasonably accurate approximation to the large-scale evolution of cosmic structures, the PM and $\texttt{COLA}$ algorithms offer a more precise treatment of nonlinear gravitational dynamics by directly integrating the equations of motion for a discretized dark matter field. The latter approach, the $\texttt{COLA}$ model, ensures that nonlinearities in structure formation, such as filamentary collapse and halo formation, are captured with greater fidelity \citep{Stopyra2024_COLA}.

A fundamental challenge in reconstructing the matter distribution from galaxy surveys is the galaxy bias problem---the fact that galaxies are not direct tracers of the total matter density field but are instead biased by astrophysical and environmental factors. \borg\ jointly infers the galaxy bias model parameters alongside the initial density field, treating bias parameters as additional latent variables in the hierarchical inference process. The galaxy distribution is modeled as an inhomogeneous Poisson process, in which the expected number density of galaxies depends nonlinearly on the underlying dark matter density field. This ensures that the inferred density field is not artificially distorted by incorrect assumptions about the biasing relation, and that uncertainties in galaxy formation processes do not propagate into the cosmological inference.

Due to the extremely high dimensionality of the problem---where the number of free parameters can exceed $10^7$, corresponding to density amplitudes at different spatial locations---\borg\ employs Hamiltonian Monte Carlo (HMC), a highly efficient gradient-based Markov Chain Monte Carlo (MCMC) technique. HMC exploits the gradient of the posterior distribution, allowing for large, correlated updates that explore parameter space far more efficiently than traditional MCMC methods. The use of HMC is critical for ensuring that \borg\ can sample from the full posterior distribution of initial conditions and galaxy bias parameters within a computationally feasible time frame.

One of the primary advantages of \borg\ is its ability to generate an ensemble of constrained realizations of the cosmic density field. Rather than producing a single deterministic reconstruction, \borg\ provides a full probabilistic description of the density field, enabling a rigorous uncertainty quantification. This approach naturally accounts for survey effects such as incomplete sky coverage, selection biases, and redshift-space distortions, all of which are marginalized over in the inference process. The resulting ensemble of density fields can then be used to study statistical properties of cosmic structure, test cosmological models, and compare with independent observational probes.

\subsection{Inference parameters and nomenclature}

The inference is performed on a cubic grid with $N$ total voxels, indexed by $i$, within a parent domain of side length $L$. The grid resolution, determined by $L/N^{\frac{1}{3}}$, sets the smallest scales that can be inferred in the initial conditions. A fiducial observer is placed at the centre of this volume, with the constrained region—where observational data directly inform the inference—surrounding this observer and extending out to a maximum radius set by the survey limits. Beyond this constrained region, the volume is unconstrained by data, containing effectively random realizations of large-scale structure. Since the local Universe is non-periodic, but the inference is performed within a periodic computational volume, this region serves to absorb boundary effects while maintaining statistical consistency.

The input dataset is divided into multiple galaxy subcatalogues, denoted by $\mathcal{C}_j$, each defined by specific luminosity, flux, and redshift cuts. These divisions account for the evolving relationship between galaxies and the underlying dark matter field, ensuring that different tracer populations are modelled separately. Each subcatalogue has its own set of bias parameters, which are inferred independently, but remain coupled through a common three-dimensional matter density field.

Selection effects are accounted for by the survey response operator $R$, which modulates the expected number of galaxies in each voxel. This operator is constructed as a three-dimensional extruded mask, combining the angular completeness of the survey with the radial selection function, the latter dictated by the flux limit of the catalogue. The inference proceeds under the assumption of a fixed cosmology, within which \borg reconstructs the large-scale structure using a Bayesian framework that jointly accounts for gravitational evolution, galaxy biasing, and survey selection effects.

\subsection{Overview of the \borg forward model}
\label{sect:borg_forward_model}

The \borg algorithm performs full field-level inference through a differentiable forward modelling approach, which maps a set of latent initial conditions to a synthetic galaxy distribution that can be compared to observations. Each iteration of the inference proceeds through the following steps:

\begin{enumerate}
    \item \textit{Sample a Gaussian white noise field}: A realization of the primordial density fluctuations is drawn from a Gaussian prior with zero mean and unit variance. This latent white noise field forms the basis for the initial conditions.

    \item \textit{Evolve forward using a gravity solver}: The sampled white noise field is modulated by the primordial power spectrum and evolved to the present day using a fast but accurate gravitational structure formation model—in our case, the \texttt{COLA} method.

    \item \textit{Transform particles to redshift space}: To match observational coordinates, peculiar velocities are used to shift dark matter particles along the line of sight, modelling redshift-space distortions (RSDs).

    \item \textit{Apply the galaxy bias model}: The evolved dark matter density field is transformed into a predicted galaxy distribution using a nonlinear, positive-definite local bias model (see \cref{sect:galaxy_bias_model}).

    \item \textit{Evaluate the likelihood and prior}: The synthetic galaxy field is compared to the observed data using a Generalized Poisson likelihood (see \cref{sect:borg_likelihood}), while the initial and final density fields are constrained by physically motivated priors (see \cref{sect:borg_prior}).

    \item \textit{Compute gradients with respect to initial conditions}: The gradients of the posterior distribution are computed with respect to the initial white noise field.

    \item \textit{Update the initial conditions}: The HMC sampler uses these gradients to propose new correlated samples of the white noise field, efficiently exploring the high-dimensional posterior distribution.
\end{enumerate}

Subsequently, the galaxy bias parameters for each subcatalogue are sampled using slice sampling \citep{Neal2000}, conditioned on the current realization of the density field, ensuring an efficient joint inference of both the phases of the white noise field and the galaxy bias parameters.

The result of this iterative procedure is a correlated Markov chain of samples from the full posterior distribution over initial conditions and bias parameters. After computing the integrated autocorrelation time of the chain (see \cref{sect:borg_performance}), we thin the chain to extract an approximately independent set of samples. We also run multiple independent chains from random starting points, both to improve sampling efficiency and to assess convergence.

This forward modelling pipeline ensures that each posterior sample is dynamically consistent, observationally constrained, and physically plausible within the \lcdm framework. In the following sections we outline the changes that were made to the \borg forward model that make up the \manticore model.

\subsection{Physics informed priors}
\label{sect:borg_prior}

In Bayesian inference, the prior encodes our assumptions about the statistical properties of the model parameters before incorporating observational data. In the context of this work, we apply prior constraints to both the initial density field—defined at a cosmic scale factor of $a \approx 10^{-3}$—and the final density field at $a = 1$. These constraints ensure that the inferred fields remain consistent with a physically plausible $\Lambda$CDM Universe. The necessity of such priors arises due to the ill-posed nature of the inverse problem: the observational process results in information loss, making a unique recovery of the initial conditions impossible based solely on the likelihood. The prior thus serves to regularize the inference, preventing the Markov chain from exploring unphysical solutions that are statistically permitted but dynamically inconsistent.

In practice, we found that traditional priors—such as enforcing only the mean and variance of the initial white noise field—were insufficient to guarantee consistency with $\Lambda$CDM expectations. In early tests, the inferred fields exhibited mild non-Gaussian features, including tilted power spectra, anisotropies in Fourier space, and excess skewness in the final density field. These behaviors arose because the field-level likelihood is highly informative, and in the absence of stronger priors, the inference prioritized likelihood maximization over physical consistency. This effect is particularly prominent when using flexible galaxy bias models that are agnostic to the underlying gravitational dynamics. As our scientific aim is to test whether observed galaxy data can be reconciled with a dynamically consistent $\Lambda$CDM model, we found it essential to explicitly encode those assumptions through physics-informed priors.

We implement three such priors: (1) a variance prior and (2) a Fourier-space power prior on the initial white noise field, and (3) moment-based priors on the final density field. The priors on the initial field encourage the statistical properties expected for a Gaussian random field—zero mean, unit variance, and a flat, isotropic power spectrum—consistent with inflationary initial conditions. The final field moments (mean, variance, skewness) are constrained using an ensemble of $\Lambda$CDM simulations that incorporate the same selection functions as the data. These refinements significantly improve the robustness of the inference and ensure that solutions remain both observationally constrained and physically meaningful within the assumed cosmology.

The following sections describe in detail the specific prior constraints applied to the initial and final density fields, along with their implementation in the inference framework.

\subsubsection{Constraining the initial white noise field}
\label{sect:prior_wnf}

In the \lcdm framework, the initial conditions for structure formation are typically prescribed by a Gaussian white noise field, which is subsequently modulated by the primordial power spectrum \citep{1985ApJS...57..241E, 1997ApJ...490L.127P}. These initial white noise fields must satisfy several key requirements. Specifically, the Gaussian assumption ensures that initial phase fluctuations are fully described by their two-point correlation function or power spectrum, with higher-order connected moments vanishing. Additionally, the fields must be invariant under translations and rotations to represent the statistical homogeneity and isotropy dictated by the cosmological principle.

In Fourier space, this implies that the corresponding power spectrum \( P(|\vec{k}|) \) is isotropic, depending only on the magnitude of the wavenumber \( |\vec{k}| \), and not on its direction. For a white noise field, the power spectrum must be flat, corresponding to \( P(|\vec{k}|) = 1 \). Note that in this context, \( P(|\vec{k}|) \) denotes the power spectrum of the latent white noise field used in the inference framework, and should not be confused with the primordial power spectrum of cosmological density perturbations. The latter typically encodes a scale-dependent modulation, e.g., \( P_{\mathrm{prim}}(k) \propto k^{n_s} \), and is applied multiplicatively to the white noise field to generate the physical initial conditions. The combination of statistical independence of the Fourier modes and a flat power spectrum ensures that the real-space covariance matrix is diagonal, as expected for a white noise field. 

To ensure physically meaningful inferences within the \lcdm framework, the prior choice for the initial white noise field is a critical component. The \borg algorithm assumes a Gaussian prior on the initial white noise field \( \mathbf{x} \):  
\begin{equation}
E_{\text{var}}(\mathbf{x}) = \frac{1}{2} \sum_{i} x_i^2,
\end{equation}  
ensuring zero mean and unit variance for the initial phase fluctuations, consistent with the statistical properties required by the \lcdm model.

This prior constrains only the total variance of the white noise field, without imposing any constraints on the power distribution in Fourier space. As such, it permits a wide range of power spectra—some of which may not be flat—so long as the total variance remains fixed to one. This flexibility can result in inferred white noise fields that are statistically inconsistent with the assumptions of isotropy and spectral flatness. In particular, it does not ensure that the off-diagonal elements of the real-space covariance matrix vanish, as required for true white noise.

To enforce the expected properties of white noise, the \manticore model incorporates an additional prior that constrains the shape of the power spectrum in Fourier space. Specifically, it penalizes deviations of the shell-averaged power from its expected flat value. As derived in \cref{sec:derive_wn_prior}, this shell-normalized power follows a Beta distribution with mean unity and a variance approximately given by:
\begin{equation}
\sigma_i^2 = 2 \left( \frac{1}{N_i} - \frac{1}{N_{\text{tot}}} \right),
\end{equation}
where \( N_i \) is the number of modes in the \( i \)-th spherical shell \( S_i \), and \( N_{\text{tot}} \) is the total number of Fourier modes in the domain.

The corresponding regularization term on the power spectrum becomes:
\begin{equation}
E_{\mathrm{ps}}(\tilde{\mathbf{x}}) = \sum_{i=1}^{N_k} \frac{\left( P(k_i) / \bar{P} - 1 \right)^2}{2 \sigma_i^2},
\end{equation}
where the shell-averaged power in shell \( S_i \) is:
\begin{equation}
P(k_i) = \frac{1}{N_i} \sum_{k \in S_i} |\tilde{\mathbf{x}}_k|^2,
\end{equation}
and the total mean power is:
\begin{equation}
\bar{P} = \frac{1}{N_{\text{tot}}} \sum_{k} |\tilde{\mathbf{x}}_k|^2.
\end{equation}

The full Hamiltonian for the prior on the initial white noise field \( \mathbf{x} \) then becomes:
\begin{equation}
H_{\text{prior}}(\mathbf{x}) = E_{\text{var}}(\mathbf{x}) + E_{\text{ps}}(\tilde{\mathbf{x}}) = \frac{1}{2} \sum_{i} x_i^2 + \frac{\lambda_{\text{ps}}}{2} \sum_{i=1}^{N_k} \frac{\left(P(k_i)/\bar{P} - 1 \right)^2}{\sigma_i^2},
\label{eq:wnf_prior_together}
\end{equation}
where \( \tilde{\mathbf{x}} \) is the Fourier transform of \( \mathbf{x} \), and \( \lambda_{\text{ps}} \) controls the relative strength of the power spectrum constraint.

In this work, we adopt a value of \( \lambda_{\text{ps}} = 100 \), chosen empirically to ensure that the shell-normalized power \( P(k_i)/\bar{P} \) remains statistically consistent with the expected variance derived in \cref{sec:derive_wn_prior}, while still allowing sufficient flexibility for the sampler to explore data-constrained regions of parameter space. We found that values of \( \lambda_{\text{ps}} \) in the range \( [50, 200] \) yielded similar results, with larger values imposing tighter conformity to white noise assumptions at the expense of reduced exploration. Our choice reflects a compromise between physical regularization and sampling efficiency.

The formulation of \cref{eq:wnf_prior_together} guides individual white noise realizations to conform both to the expected variance of a standard Gaussian field and to the flat, isotropic power spectrum characteristic of white noise in \lcdm cosmology.
 
\subsubsection{Constraining the moments of the final density field}
\label{sect:prior_final_field}

In addition to constraining the frequency-space properties of the initial white noise field, we impose priors on the statistical moments of the final density field. For each galaxy subcatalogue, we require that the inferred dark matter density field within its selection window exhibits statistical properties that are consistent with expectations from the \lcdm model. Specifically, we constrain the weighted mean, variance, and skewness of the density field within the selected voxels to match the distributions derived from an ensemble of random \lcdm simulations, subject to the same survey selection effects. These constraints ensure that the inferred density field remains statistically representative of a \lcdm realization while still accommodating the constraints imposed by the observed galaxy distribution.

These one-point moments of the final density field are computed within the selection window of each galaxy subcatalogue, \( \mathcal{C}_j \), and take the following weighted forms:

\begin{align}
\text{Weighted mean:} & \quad \bar{\delta}_{\mathcal{C}_j} = \frac{\sum\limits_{i \in \mathcal{C}_j} R_i \delta_i}{\sum\limits_{i \in \mathcal{C}_j} R_i}, \\
\text{Weighted variance:} & \quad \tilde{s}^2_{\mathcal{C}_j} = \frac{\sum\limits_{i \in \mathcal{C}_j} R_i (\delta_i - \bar{\delta}_{\mathcal{C}_j})^2}{\sum\limits_{i \in \mathcal{C}_j} R_i}, \\
\text{Weighted skewness:} & \quad \tilde{\theta}_{\mathcal{C}_j} = \frac{\sum\limits_{i \in \mathcal{C}_j} R_i \left( \frac{\delta_i - \bar{\delta}_{\mathcal{C}_j}}{\tilde{s}_{\mathcal{C}_j}}\right)^3 }{\sum\limits_{i \in \mathcal{C}_j} R_i}, \\
\text{Weighted kurtosis:} & \quad \tilde{\Omega}_{\mathcal{C}_j} = \frac{\sum\limits_{i \in \mathcal{C}_j} R_i \left( \frac{\delta_i - \bar{\delta}_{\mathcal{C}_j}}{\tilde{s}_{\mathcal{C}_j}}\right)^4 }{\sum\limits_{i \in \mathcal{C}_j} R_i} - 3.
\end{align}

\begin{figure}
    \includegraphics[width=\columnwidth]{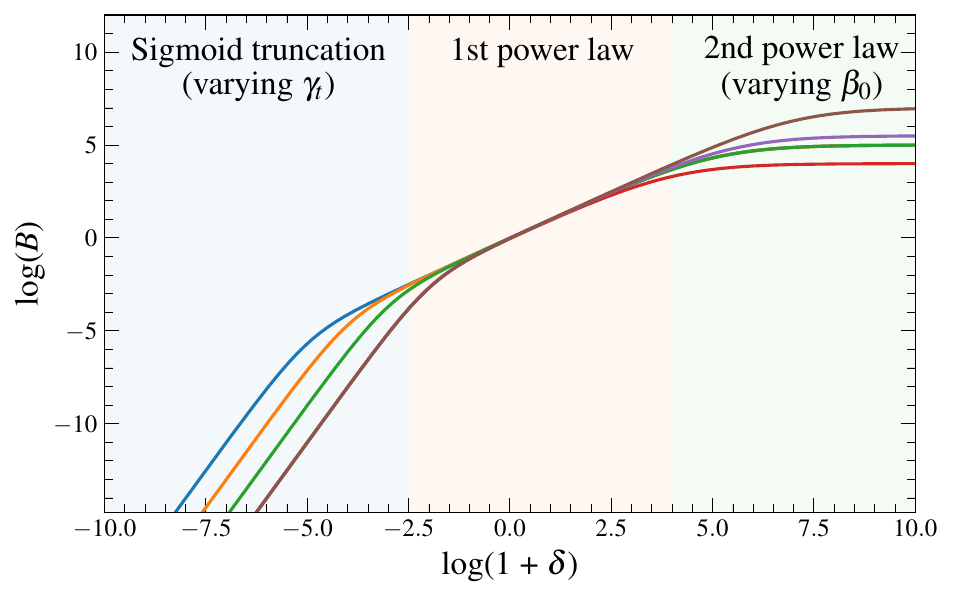}
    \caption{A toy example illustrating the flexibility of the three-component galaxy bias model described in \cref{sect:galaxy_bias_model}. The figure depicts the predicted galaxy number density ($B$) as a function of the dark matter density ($1+\delta$), demonstrating how the model responds to variations in two key parameters: $\gamma_t$, which controls the transition between the sigmoid truncation and the first power-law regime, and $\beta_0$, which governs the transition between the first and second power-law components.}
    \label{fig:toy_bias_demo}
\end{figure}

\noindent
Here, \( R_i \) is the survey response operator, which encodes selection effects by combining the angular survey mask and the radial selection function dictated by the survey flux limit. The index \( i \) runs over the voxels within the selection window of galaxy subcatalogue \( \mathcal{C}_j \).

To construct a physically motivated prior, we compute the expected moments of the final density field from an ensemble of 1000 random \lcdm simulations using the same \texttt{COLA} gravity solver that is used during the \borg inference. For each realization, the survey response operator of each galaxy subcatalogue is applied to the simulated final density field, and the cumulative moments are computed following the definitions above.

A Gaussian prior is then imposed on the weighted mean, variance, and skewness of the final density field for each galaxy subcatalogue. The weighted kurtosis is retained as a validation metric to assess whether higher-order moments of the inferred density field remain consistent with \lcdm expectations. The prior likelihood takes the form:

\begin{equation}
E_{\mathrm{mom}}(X_{\mathcal{C}_j}) = \frac{1}{2} \frac{\left( X_{\mathcal{C}_j} - \mu \right)^2}{(\gamma \sigma)^2}, 
\label{eq:prior_moment_likelihood}
\end{equation}

\noindent
where \( X_{\mathcal{C}_j} \) represents one of the moments \( \bar{\delta}_{\mathcal{C}_j} \), \( \tilde{s}^2_{\mathcal{C}_j} \), or \( \tilde{\theta}_{\mathcal{C}_j} \). The parameters \( \mu \) and \( \sigma \) denote the mean and standard deviation of the respective moment, as determined from the 1000 \lcdm realizations. The hyperparameter \( \gamma \) regulates the strictness of the prior.

We adopt a value of \( \gamma = 0.5 \), chosen empirically to balance the strength of the regularization. This value ensures that the inferred one-point statistics of the final density field—specifically the weighted mean, variance, and skewness—remain within approximately \( 1\sigma \) of the expectations derived from our ensemble of \lcdm simulations. Smaller values of \( \gamma \) enforce tighter conformity to the \lcdm moment distributions, but risk suppressing legitimate fluctuations supported by the data, while larger values permit greater deviations at the cost of potentially reduced physical consistency. We found \( \gamma = 0.5 \) to be a robust compromise across all subcatalogues, effectively regularizing the inference without over-constraining it.

The total Hamiltonian for the moment prior is then formulated as:

\begin{equation}
    H_{\mathrm{mom}} = \sum\limits_{j \in N_{\mathrm{cat}}} \left[ E_{\mathrm{mom}}(\bar \delta_{\mathcal{C}_j}) + E_{\mathrm{mom}}(\tilde{s}^2_{\mathcal{C}_j}) + E_{\mathrm{mom}}(\tilde{\theta}_{\mathcal{C}_j}) \right],
\end{equation}

\noindent
where \( N_{\mathrm{cat}} \) is the number of galaxy subcatalogues.

\subsection{The galaxy bias model}
\label{sect:galaxy_bias_model}

The transformation from the dark matter density field to the observed galaxy field is governed by a nonlinear, positive-definite local bias model. Each galaxy subcatalogue is assigned an independent set of free parameters within this model to account for variations in tracer properties across different luminosity, flux and redshift cuts. In this section, we describe the procedure for estimating the expected galaxy counts within a single voxel (denoted \( i \)) for a specific galaxy subcatalogue (denoted \( \mathcal{C}_j \)). To simplify notation, we temporarily omit the \( i \) and \( \mathcal{C}_j \) subscripts.

\subsubsection{A sigmoid-truncated double power-law galaxy bias model}

The galaxy bias model introduced in this work extends previous formulations used in Aquila studies \citep[e.g.,][]{Jasche2019, Stopyra2021}, originally based on \citet{Neyrinck2014}. We employ a Sigmoid-Truncated Double Power-Law (STDP) bias model, which refines the existing truncated power-law model by introducing an additional power-law regime at high densities and replacing the exponential cutoff with a sigmoid truncation. The key motivation remains the same: galaxy formation is suppressed in low-density environments while a power-law relation is maintained over intermediate density regimes. The introduction of a second power-law component provides additional flexibility in high-density regions, allowing for potential effects such as feedback-driven quenching of galaxy formation \citep[e.g.,][]{Behroozi2013}.

The model consists of three multiplicative components that depend only on the local dark matter density, $\rho = 1 + \delta$, and a set of free parameters. The first component, a sigmoid truncation, regulates galaxy formation at low densities:

\begin{equation}
    \Theta_{\mathrm{sig}}(\rho, \gamma_t, \gamma_s) = \frac{1}{1+\exp[(\gamma_t - \ln(\rho)) / \gamma_s]},
\end{equation}

\noindent where \( \gamma_t \) defines the transition point between the truncation and the first power-law regime, while \( \gamma_s \) controls the steepness of the transition.

The second component is a primary power-law, representing the dominant relationship between the dark matter density and galaxy formation:

\begin{equation}
    \Theta_{\mathrm{1st}}(\rho, \alpha) = \rho^{\alpha},
\end{equation}

\noindent where \( \alpha \) is the power-law exponent.

The final component introduces a secondary power-law, allowing for a gradual transition at higher densities:

\begin{equation}
    \Theta_{\mathrm{2nd}}(\rho, \beta, \beta_0) = \left( 1+\exp[\beta (\ln(\rho) - \beta_0)] \right)^{-\beta},
\end{equation}

\noindent where \( \beta \) defines the slope of the second power-law component, and \( \beta_0 \) sets the transition scale between the two power-law regimes. If \( \beta = 0 \), the model reduces to a single power-law.

The expected galaxy count per voxel is then given by the product of these three components:

\begin{equation}
    B(\rho, \gamma_t, \gamma_s, \alpha, \beta, \beta_0) =  \Theta_{\mathrm{sig}}(\rho, \gamma_t, \gamma_s) \Theta_{\mathrm{1st}}(\rho, \alpha) \Theta_{\mathrm{2nd}}(\rho, \beta, \beta_0).
    \label{eq:galaxy_bias}
\end{equation}

\noindent The bias function is deliberately constructed without an explicit normalization factor (see below).

Figure \ref{fig:toy_bias_demo} illustrates how different parameter values affect the bias function, demonstrating its flexibility across various density regimes. Since the bias parameters are not known \textit{a priori}, they are inferred jointly with the initial conditions during the sampling process.

\subsubsection{Normalizing the galaxy bias to the ergodic mean}

A nonlinear transformation of the matter density field through a bias function does not inherently preserve the ergodic mean of the galaxy field. Specifically, applying a general bias function can shift the mean galaxy overdensity, violating the requirement that:

\begin{equation}
    \langle \delta_g \rangle = 0.
\end{equation}

This condition is satisfied only if the ergodic mean of the biased galaxy field equals unity:

\begin{equation}
    \langle B'(\delta_m) \rangle = 1.
\end{equation}

To enforce this condition, we normalize the bias function by dividing by its ergodic mean:

\begin{equation}
    B'(\delta_m) = \frac{B(\delta_m)}{\langle B(\delta_m) \rangle}.
\end{equation}

The ergodic mean is estimated at each inference step by averaging the bias function over the full parent domain:

\begin{equation}
    \langle B \rangle = \frac{1}{N_{\mathrm{vox}}} \sum_{i=0}^{N_{\mathrm{vox}}-1} B_i,
\end{equation}

\noindent where \( N_{\mathrm{vox}} \) is the total number of voxels in the computational domain.

Restoring the voxel and subcatalogue indices, the expected number of galaxies in voxel \( i \) for galaxy subcatalogue \( \mathcal{C}_j \) is then given by:

\begin{equation}
    \lambda_i^{\mathcal{C}_j} = \bar{N}^{\mathcal{C}_j} R_i^{\mathcal{C}_j} \frac{B_i^{\mathcal{C}_j}}{\langle B^{\mathcal{C}_j} \rangle},
    \label{eq:poisson_intensity}
\end{equation}

\noindent where \( \bar{N}^{\mathcal{C}_j} \) is an additional free parameter representing the mean number of galaxies in the catalogue \( \mathcal{C}_j \), and \( R_i^{\mathcal{C}_j} \) is the survey response operator at voxel \( i \), encoding selection effects.

Unlike previous formulations, \( \bar{N}^{\mathcal{C}_j} \) is explicitly decoupled\footnote{In reality $\bar N$ is never fully decoupled from the bias due to the survey mask.} from the bias function, ensuring that the sampling of \( \bar{N}^{\mathcal{C}_j} \) remains independent of the bias parameters. This effectively removes degeneracies between the amplitude of the density field and any global normalization parameter, improving the stability and convergence of the inference.

\subsection{A generalized Poisson likelihood}
\label{sect:borg_likelihood}

The galaxy distribution is commonly modelled as a discrete sampling of the continuous underlying dark matter field. Traditionally, the assumption is that the galaxy density field follows a Poisson sampling process within each voxel \citep[for a comprehensive review of galaxy bias models and the foundations of these assumptions, see][and references therein]{Desjacques2018}. In a standard Poisson distribution, the mean and variance are equal. However, several studies have observed super-Poissonian behavior in the scatter of galaxy counts, particularly in high-density regions, where the variance exceeds the mean \citep[e.g.,][]{Kitaura2014, Neyrinck2014, Ata2015, Gruen2018, Britt2024}.

To account for this overdispersion, we adopt a Generalized Poisson likelihood, which introduces an additional free parameter to regulate the variance independently of the mean. The likelihood for a single voxel, given a specific galaxy subcatalogue (omitting explicit \( i \) and \( \mathcal{C}_j \) indices for clarity), is given by:

\begin{equation}
    \Phi (N|\lambda, b) = \frac{\lambda (1-b)}{N!} \left[ \lambda (1-b) + N b \right]^{N-1} \exp \left[ -\lambda (1-b) - N b \right],
\end{equation}

\noindent where \( \lambda \) is the Poisson intensity from \cref{eq:poisson_intensity}, \( N \) is the observed number of galaxies, and \( b \in [0,1] \) is the overdispersion parameter, which governs the deviation from the standard Poisson variance. The mean of the distribution remains:

\begin{equation}
    \mu = \lambda,
\end{equation}

\noindent while the variance is modified to:

\begin{equation}
    \sigma^2 = \frac{\lambda}{(1-b)^2}.
\end{equation}

When \( b = 0 \), the standard Poisson case is recovered, in which the mean and variance are equal. Positive values of \( b \) allow for an increased variance, accommodating the possibility that galaxy counts exhibit larger fluctuations than a pure Poisson process would predict.

Observational and simulation studies suggest that the level of overdispersion is not spatially uniform but instead varies with local density and halo mass. This trend has been identified in previous analyses and is further supported by our preliminary investigation of galaxy count scatter as a function of dark matter density in the cosmological hydrodynamical \flamingo simulation \citep{Schaye2023}. Our findings indicate that regions of higher density systematically exhibit greater variance in galaxy counts. To reflect this behavior, we restrict \( b \) to positive values, ensuring that we only model overdispersion and do not introduce artificial underdispersion.

To capture the density dependence of overdispersion, we parametrize \( b \) as a function of the local dark matter density $\rho = 1 + \delta$ using a power-law relation:

\begin{equation} 
    b(\rho, b_{\alpha}, b_{\beta}) = b_{\alpha} \rho^{b_{\beta}},
\end{equation}

\noindent where \( b_{\alpha} \) sets the normalization and \( b_{\beta} \) determines the slope of the power-law scaling. If \( b_{\beta} = 0 \), the overdispersion remains constant across all densities. As with the free parameters of the galaxy bias model, \( b_{\alpha} \) and \( b_{\beta} \) are unknown \textit{a priori} and are inferred alongside the initial density field.

Restoring the explicit voxel and galaxy subcatalogue indices, the full likelihood function is expressed as:

\begin{equation}
    \mathcal{L} = \sum\limits_{j \in N_\mathrm{cat}} \sum\limits_{i \in \mathcal{C}_j} \Phi \left( N_i^{\mathcal{C}_j} \mid \lambda_i^{\mathcal{C}_j}, b_i^{\mathcal{C}_j} \right).
\end{equation}

\noindent This expression sums the likelihood contributions over all voxels within the domain of each galaxy subcatalogue and across all subcatalogues. Here, \( N_i^{\mathcal{C}_j} \) represents the observed galaxy counts in voxel \( i \) for galaxy subcatalogue \( \mathcal{C}_j \), while \( \lambda_i^{\mathcal{C}_j} \) and \( b_i^{\mathcal{C}_j} \) correspond to the expected galaxy counts and dispersion level, respectively, as determined by the current state of the galaxy bias parameters.

By incorporating a Generalized Poisson likelihood, we introduce a more flexible and physically motivated description of galaxy number counts, improving the inference process by allowing for variations in the count variance. This accounts for deviations from idealized Poisson statistics that arise due to complex astrophysical processes such as galaxy formation efficiency, feedback, and environmental effects, providing a more realistic statistical model for the observed data.

The \manticore model builds on the core \borg framework by incorporating a more flexible galaxy bias prescription, a generalized Poisson likelihood to account for overdispersion in galaxy counts, and physics-informed priors that enforce consistency with \lcdm expectations in both the initial and final density fields. Together, these improvements enhance the robustness, physical interpretability, and statistical fidelity of the inference, enabling more accurate reconstructions of the large-scale structure of the local Universe.
\section{Constraints for White Noise}
\label{sec:derive_wn_prior}

To construct physically meaningful initial conditions, the \borg algorithm assumes that the input white noise field is a realization of a statistically homogeneous and isotropic Gaussian random field. This implies that the field’s power spectrum in Fourier space should be both flat and isotropic. Under the assumption of unit variance, the power spectrum of a statistically isotropic white noise field is flat, with
\begin{equation}
P(|\vec{k}|) = 1 \quad \text{for all } |\vec{k}|,
\end{equation}
where \( P(|\vec{k}|) \) denotes the power spectrum as a function of the wavenumber magnitude. To ensure statistical isotropy, the power spectrum must be invariant under rotations in Fourier space—that is, it must be uniform across spherical shells (or frequency bins) of constant \( |\vec{k}| \).

For a discrete field \( \mathbf{x} \in \mathbb{R}^N \), we denote its Fourier transform by \( \tilde{\mathbf{x}} \), and define the shell-averaged normalized power \( m_i \) in the \( i \)-th shell \( S_i \) as:
\begin{equation}
m_i = \frac{N_{\text{tot}}}{N_i} \cdot \frac{\sum_{k \in S_i} |\tilde{\mathbf{x}}_k|^2}{\sum_{k \in S} |\tilde{\mathbf{x}}_k|^2},
\end{equation}
where \( N_i = |S_i| \) is the number of Fourier modes in shell \( S_i \), and \( N_{\text{tot}} = |S| \) is the total number of Fourier modes. This definition ensures that a white noise field satisfies \( \mathbb{E}[m_i] = 1 \) for all \( i \), and that deviations from this condition indicate anisotropy or spectral tilt. In what follows, we derive the expected distribution and variance of \( m_i \), which will motivate the prior imposed on the power spectrum shape in \cref{sect:prior_wnf}.

We begin by defining:
\begin{equation}
A = \sum_{k \in S_i} |\tilde{\mathbf{x}}_k|^2, \quad B = \sum_{k \in S} |\tilde{\mathbf{x}}_k|^2,
\end{equation}
where \( A \) is the total power in shell \( S_i \), and \( B \) is the total power over all modes. Then,
\begin{equation}
m_i = \frac{N_{\text{tot}}}{N_i} \cdot \frac{A}{B}.
\end{equation}

Because \( \tilde{\mathbf{x}}_k \) are independent standard Gaussian variables for white noise fields, \( A \) and \( B \) are sums of squared Gaussian variables and therefore follow Gamma distributions. Their ratio \( A/B \), which involves the ratio of two independent Gamma-distributed variables with the same scale parameter, follows a Beta distribution:
\begin{equation}
\frac{A}{B} \sim \text{Beta}\left(\frac{N_i}{2}, \frac{N_{\text{tot}} - N_i}{2}\right).
\end{equation}

The expectation of a Beta-distributed variable \( X \sim \text{Beta}(\alpha, \beta) \) is:
\begin{equation}
\mathbb{E}[X] = \frac{\alpha}{\alpha + \beta}.
\end{equation}
Applying this to our case:
\begin{equation}
\mathbb{E} \left[ \frac{A}{B} \right] = \frac{N_i}{N_{\text{tot}}} \quad \Rightarrow \quad \mathbb{E}[m_i] = \frac{N_{\text{tot}}}{N_i} \cdot \frac{N_i}{N_{\text{tot}}} = 1.
\end{equation}

To approximate the prior as a Gaussian, we compute the variance of \( m_i \). The variance of a Beta-distributed variable is:
\begin{equation}
\text{Var}(X) = \frac{\alpha \beta}{(\alpha + \beta)^2 (\alpha + \beta + 1)}.
\end{equation}
In our case, this becomes:
\begin{equation}
\text{Var} \left( \frac{A}{B} \right) = \frac{\frac{N_i}{2} \cdot \frac{N_{\text{tot}} - N_i}{2}}{\left( \frac{N_{\text{tot}}}{2} \right)^2 \left( \frac{N_{\text{tot}} + 2}{2} \right)} = \frac{N_i (N_{\text{tot}} - N_i)}{N_{\text{tot}}^2 (N_{\text{tot}} + 2)}.
\end{equation}
For large \( N_{\text{tot}} \), we approximate:
\begin{equation}
\text{Var} \left( \frac{A}{B} \right) \approx \frac{N_i (N_{\text{tot}} - N_i)}{N_{\text{tot}}^3}.
\end{equation}
Therefore, the variance of \( m_i \) becomes:
\begin{equation}
\text{Var}(m_i) = \left( \frac{N_{\text{tot}}}{N_i} \right)^2 \cdot \text{Var}\left( \frac{A}{B} \right) \approx \frac{N_{\text{tot}}^2}{N_i^2} \cdot \frac{N_i (N_{\text{tot}} - N_i)}{N_{\text{tot}}^3} = \frac{N_{\text{tot}} - N_i}{N_i N_{\text{tot}}}.
\end{equation}
Rewriting this in a more interpretable form:
\begin{equation}
\text{Var}(m_i) \approx 2 \left( \frac{1}{N_i} - \frac{1}{N_{\text{tot}}} \right).
\end{equation}

We thus conclude that for a statistically isotropic white noise field, the normalized shell power \( m_i \) follows a Beta distribution with mean unity and variance approximately:
\[
\text{Var}(m_i) \approx 2 \left( \frac{1}{N_i} - \frac{1}{N_{\text{tot}}} \right),
\]
in the large-\( N_{\text{tot}} \) limit. This derivation justifies the use of a Gaussian prior on the quantity \( m_i = P(|\vec{k}_i|)/\bar{P} \), which appears in the power spectrum regularization term described in \cref{sect:prior_wnf}. There, we penalize deviations of \( m_i \) from unity across all Fourier shells, enforcing the flatness and isotropy expected for white noise fields. The variance derived here is used to weight the contribution from each shell in the total prior energy \( E_{\mathrm{ps}}(\mathbf{x}) \). This prior forms a key part of the \manticore model, ensuring that individual initial condition samples are consistent with the statistical properties of the \lcdm cosmology.

\section{A \lcdm-like Parent Volume (cont.)}
\label{sect:parent_lcdm_cont}

This appendix presents the detailed results of the \lcdm validation tests introduced in \cref{sect:lcdm_parent}. These include comparisons of the halo mass function (HMF) and the Gaussianity of the initial white noise fields.

\begin{figure}
    \includegraphics[width=\columnwidth]{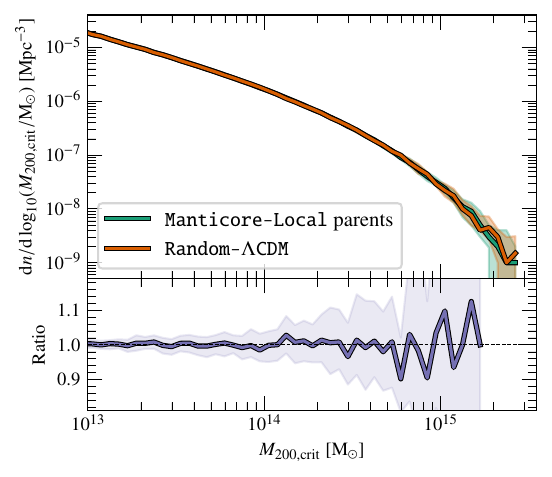}
    \caption{Comparison of the halo mass function (HMF) at \(z=0\) between the posterior resimulations and the control \lcdm simulations. The upper panel shows the median HMFs, with shaded regions indicating the \tentoninety percentile ranges across realizations. The lower panel displays the ratio of posterior to control counts, showing consistency with unity across the full halo mass range. This agreement confirms that the inferred volumes reproduce the expected mass hierarchy of structure formation in \lcdm cosmology.}
    \label{fig:parent_halo_mass}
\end{figure}

\Cref{fig:parent_halo_mass} compares the HMF of the posterior resimulations to that of a control ensemble of ten unconstrained \lcdm simulations. The distributions exhibit strong agreement across the full mass range. The ratio remains consistent with unity to within a few percent for \masscrit~$< 10^{14}$~\msol, confirming that halo abundances in the inferred volume follow the expected structure formation hierarchy of \lcdm cosmology.

\begin{figure}
    \includegraphics[width=\columnwidth]{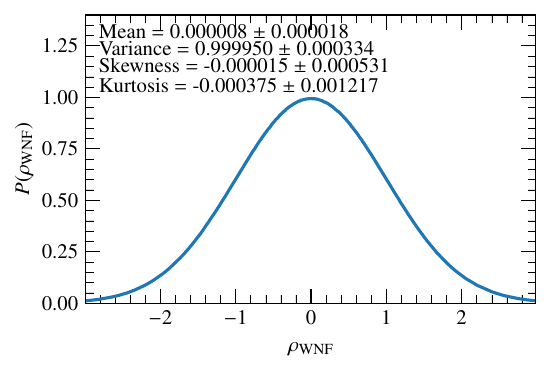}
    \caption{Probability density function (PDF) of white noise field values from the posterior samples. The solid line shows the median PDF, with the shaded region spanning the \tentoninety percentile range across realizations (hidden behind solid line). The mean, variance, skewness, and kurtosis are annotated with uncertainties and are all consistent with Gaussian expectations to within \(1\sigma\).}
    \label{fig:1d_pdf}
\end{figure}

\begin{figure}
    \includegraphics[width=\columnwidth]{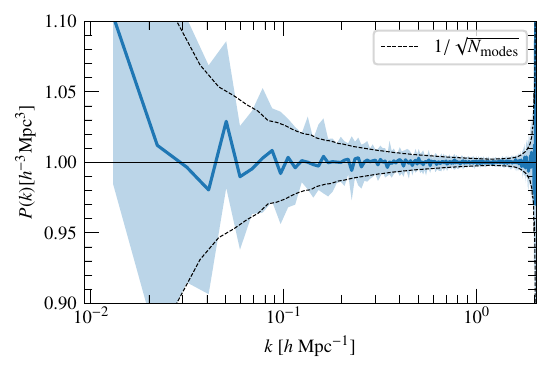}
    \caption{Normalized power spectrum of the white noise fields across the posterior ensemble. The solid line shows the median, and the shaded region indicates the \tentoninety percentile spread. The dashed line marks the expected \(1/\sqrt{N}\) Poisson noise level, where \(N\) is the number of modes in each \(k\)-bin. The posterior samples remain within this scatter, confirming the flat power spectrum expected for Gaussian white noise.}
    \label{fig:wnf_ps}
\end{figure}

\begin{figure}
    \includegraphics[width=\columnwidth]{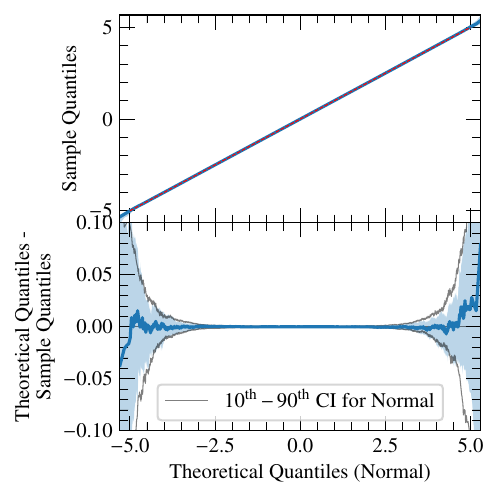}
    \caption{Quantile-quantile (Q-Q) plot comparing the distribution of white noise field values in the posterior ensemble to a standard normal distribution. The upper panel shows the quantiles, with the diagonal line representing perfect agreement. The lower panel displays residuals between the posterior and theoretical quantiles. The shaded region shows the \tentoninety percentile range of the posterior, while black lines outline the expected scatter from an equal number of random normal draws. The residuals remain within expectations, supporting consistency with a Gaussian distribution.}
    \label{fig:wnf_qq}
\end{figure}

To validate the Gaussianity of the inferred initial conditions, we compute the one-point PDF of the white noise fields (\cref{fig:1d_pdf}), their normalized power spectra (\cref{fig:wnf_ps}), and Quantile-Quantile plots (\cref{fig:wnf_qq}). All of the first four statistical moments—mean, variance, skewness, and kurtosis—lie within \(1\sigma\) of the values expected for a Gaussian random field. The power spectra are flat across all realizations, with fluctuations consistent with Poisson statistics. The Q-Q residuals remain well within the expected range from normal samples.

Together, these diagnostics confirm that the parent volumes in our posterior resimulations satisfy the statistical assumptions of \lcdm cosmology in both their initial and final states. These results validate the robustness of our inference pipeline and support the physical plausibility of the reconstructed local Universe.
\textbf{}
\section{\textsc{BORG} Performance and MCMC Chain Convergence}
\label{sect:borg_performance}

\begin{figure}
    \includegraphics[width=\columnwidth]{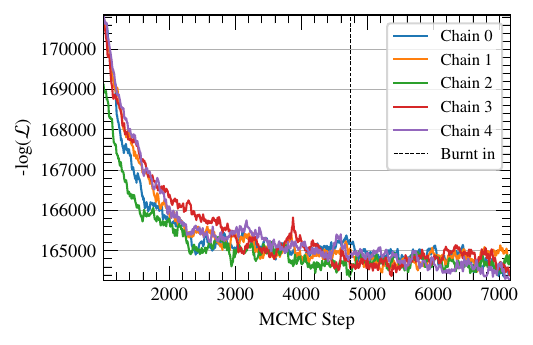}
    \caption{Trace plot of the negative log-likelihood from MCMC step 1000 onward (i.e., after the \textsc{COLA} solver was activated) for each of the five independent MCMC chains. The vertical dashed line at step 4750 marks the threshold used to define the start of post-burn-in sampling. The likelihood is stable and flat beyond this point, indicating the chain has reached equilibrium.}
    \label{fig:likelihood_trace}
\end{figure}

\begin{figure}
    \centering
    \begin{subfigure}[b]{0.45\textwidth}
        \centering
        \includegraphics[width=\textwidth]{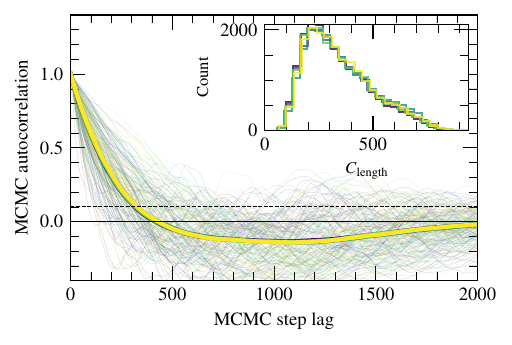}
        \caption{Autocorrelation for final-density voxels}
        \label{fig:corr_length_density}
    \end{subfigure}
    \hfill
    \begin{subfigure}[b]{0.45\textwidth}
        \centering
        \includegraphics[width=\textwidth]{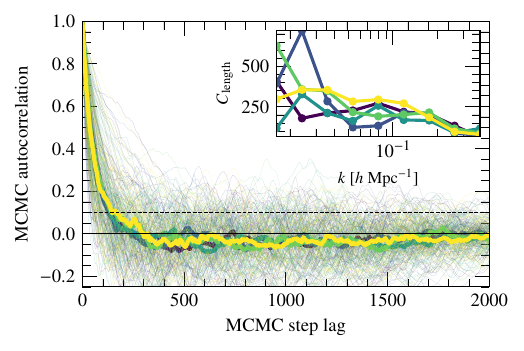}
        \caption{Autocorrelation for power spectrum modes}
        \label{fig:corr_length_ps}
    \end{subfigure}
    \caption{Autocorrelation length diagnostics for the final-density field. Each colour represents one of the five MCMC chains. In both panels, faded lines show the autocorrelation functions for individual modes or voxels, while bold curves trace the median correlation length per chain. \textbf{Top:} Distribution of correlation lengths across 20,000 voxels within the constrained region ($R=1$). The median is $\approx 250$ steps. \textbf{Bottom:} Correlation length as a function of wavenumber $k$ for power spectrum modes. Large scales ($k < 0.05\,h\,\mathrm{Mpc}^{-1}$) decorrelate more slowly, with correlation lengths reaching up to $\approx 600$ steps.}

    \label{fig:joint}
\end{figure}

\begin{figure*}
    \includegraphics[width=\textwidth]{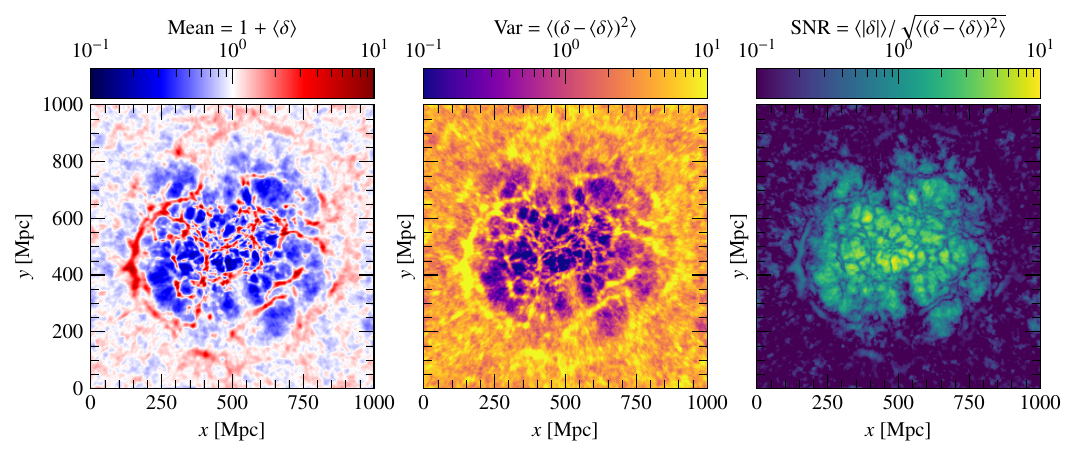}
    \caption{Mean, variance, and signal-to-noise ratio (SNR) of the final-density field across posterior samples. Shown is a 20 Mpc slice across the full $L = 1$~Gpc parent domain. Coherent structure with low variance and high SNR is visible within the constrained region ($R \lesssim 200$ Mpc), while the unconstrained outer volume tends toward a prior-dominated mean of zero.}
    \label{fig:field_variance}
\end{figure*}

\begin{figure}
    \includegraphics[width=\columnwidth]{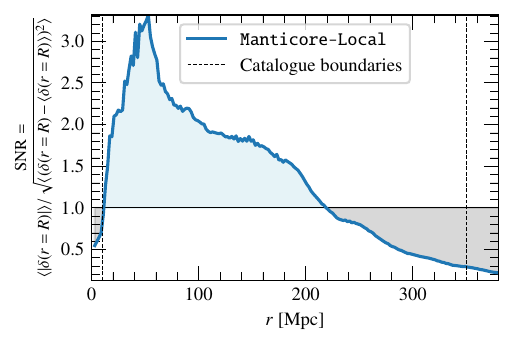}
    \caption{Mean signal-to-noise ratio (SNR) of the inferred final-density field in radial shells centred on the observer. The vertical dashed lines denote the bright ($R=10$~Mpc) and faint ($R=350$~Mpc) magnitude limits of the input galaxy catalogues, respectively. The region between 10 and 200~Mpc maintains SNR above unity and defines the fiducial analysis volume.}
    \label{fig:snr_vs_r}
\end{figure}

The \textsc{BORG} algorithm uses Markov Chain Monte Carlo (MCMC) sampling to explore the posterior distribution of initial conditions in the local Universe. To ensure robust convergence and broad posterior coverage, we employed five independent MCMC chains, each initialized with a distinct random seed. This approach mitigates the risk of chains becoming trapped in local maxima and provides a principled measure of convergence across the ensemble.

Each chain was run for 7,150 steps, with the first 4,750 steps allocated to a structured burn-in phase. During this period, two key components were progressively "warmed up" to allow for stable and efficient convergence. First, the accuracy of the \textsc{COLA} gravity solver was gradually increased by raising the \texttt{forcesampling} parameter from 2 to its final value of 4. This parameter controls the number of force evaluations per time step and directly affects the resolution of nonlinear structure formation. We adopt a final value of 4 based on the convergence studies of \citet{Stopyra2024_COLA}.

Second, the prior regularization on the final density field (see \cref{sect:prior_final_field}) was incrementally tightened by reducing the hyperparameter $\gamma$ (see \cref{eq:prior_moment_likelihood}), from an initial loose setting of 2.0 to its fiducial value of 0.5. This gradual reduction allows the sampler to first explore broad solutions before enforcing closer statistical agreement with the \lcdm prior moments. The hyperparameter controlling the initial white noise prior, $\lambda_{\mathrm{ps}}$, was fixed at 100 throughout sampling.

The transition to full settings—$\texttt{forcesampling} = 4$ and $\gamma = 0.5$—was completed by step 3,000. The remaining 1,750 burn-in steps allowed the system to stabilize under the final target distribution, before post-burn-in analysis commenced at step 4,750. Each chain was executed on 512 computing cores (64 tasks with 8 threads per task), with a total runtime of approximately 750k CPU hours per chain.

\cref{fig:likelihood_trace} shows the trace of the negative log-likelihood for MCMC steps 1000 through 7150. In the early phase, the likelihood steadily increases (i.e., the negative log-likelihood decreases) as the sampler moves from an initial state of poor model fit toward regions of higher posterior probability. After approximately step 4750, the curve transitions to a regime of stochastic but bounded fluctuations, indicative of the chain entering the typical set of the target distribution. This stabilization is characteristic of convergence in high-dimensional HMC sampling, where the sampler explores a high-probability shell rather than continuing toward the peak of the distribution. We therefore conservatively adopt step 4750 as the end of the burn-in period and the beginning of post-burn-in analysis.

To quantitatively assess convergence after burn-in, we compute the Gelman--Rubin statistic (\(\hat{R}\)), which compares within-chain and between-chain variances as a measure of sampling consistency across independent chains. For the final 2,400 steps (from step 4750 to 7150), corresponding to the post-burn-in phase identified via likelihood stabilization, we find \(\hat{R} = 1.0573\) for the negative log-likelihood. This satisfies the conventional convergence criterion of \(\hat{R} < 1.1\) \citep{Gelman1992}, supporting the interpretation that all chains are sampling from the same posterior distribution and have reached a common stationary regime.

We assess the sampling efficiency of the post–burn-in chains via the autocorrelation length. Because HMC introduces correlations between successive samples, the number of effectively independent samples is determined by how quickly the chain decorrelates. \Cref{fig:joint} displays autocorrelation statistics for individual voxels within the fully constrained region ($R=1$) of the final density field, as well as for individual power spectrum modes. In both panels, we show the autocorrelation curves for individual voxels or modes (faded lines) alongside the median trend for each chain.

For density voxels, the median autocorrelation length is approximately 250 steps, with a 10th–90th percentile range of 150–400 steps. Power spectrum modes decorrelate at scale-dependent rates: smaller-scale modes ($k > 0.1\,h\,\mathrm{Mpc}^{-1}$) decorrelate rapidly, while larger-scale modes ($k < 0.05\,h\,\mathrm{Mpc}^{-1}$) exhibit longer correlation lengths of up to $\sim 600$ steps.

To complement these diagnostics, we compute the effective sample size (ESS) using the standard estimator $1 / \left(1 + 2 \sum_k |\rho_k|\right)$, where $\rho_k$ is the autocorrelation at lag $k$. For power spectrum modes, we find a median ESS of 20 (with a 10th–90th percentile range of 7–40), and for density voxels, a median ESS of 8 (with a 10th–90th percentile range of 5–14). These values are consistent across chains and confirm that our adopted thinning interval of 250 steps—yielding 10 effective samples per chain, or 50 in total—strikes a reasonable balance between sampling efficiency and posterior coverage. The similarity of median autocorrelation trends across chains further supports our assessment of convergence.

\cref{fig:field_variance} illustrates the posterior mean, variance, and corresponding SNR of the final density field across the entire volume within a 20~Mpc slice. Within the constrained region—where data directly inform the inference—the posterior mean density field displays coherent structure, while the variance remains suppressed due to the constraining power of the observations. In contrast, outside this region, where observational coverage is lacking, the mean field averages toward zero and the variance increases, reflecting the field's reversion to the unconstrained prior. This transition is also clearly visible in the radial SNR profile shown in \cref{fig:snr_vs_r}, where the SNR remains above unity within the range \( R = 10 \)--200 Mpc, peaking at values around 3 near \( R \approx 50 \) Mpc.

The radial extent of the constrained region is determined by the selection choices inherent to the input galaxy catalogs, which arise from the interplay between apparent magnitude limits (imposed by survey flux thresholds) and absolute magnitude cuts (used to define uniform tracer populations). These constraints introduce a bright-end cutoff at small distances: for galaxies brighter than the apparent magnitude limit, the corresponding minimum observable distance is approximately \( R = 10 \) Mpc. Conversely, a faint-end cutoff limits the maximum distance to which galaxies can be detected, as intrinsically faint objects fall below the flux threshold at large radii. While the formal selection function extends out to \( R \approx 350 \) Mpc, the combination of decreasing tracer density and growing shot noise causes the signal-to-noise ratio to fall below unity beyond \( R \approx 200 \) Mpc. This transition indicates that the posterior becomes increasingly dominated by the prior at larger radii. We therefore conservatively define the radial range \( R = 10 \)--200 Mpc as our fiducial constrained analysis volume, within which the inference remains reliably data-informed.

In summary, the \textsc{BORG} inference presented here achieves strong convergence, reasonable autocorrelation lengths, and high signal-to-noise within the constrained volume. The use of multiple chains, adaptive burn-in, and careful sampling diagnostics ensures the reliability of our posterior ensemble, forming a robust foundation for the analyses presented in this work.

\section{Inference convergence and \manticoremini}
\label{sect:resolution_convergence}

\begin{figure}
    \centering
    \includegraphics[width=\columnwidth]{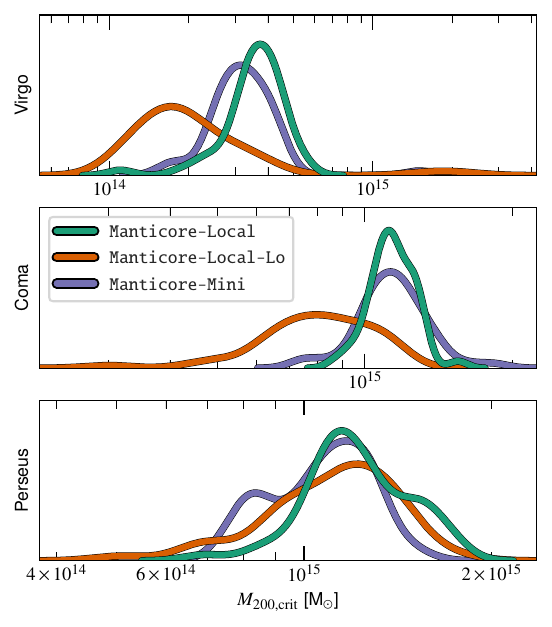}
    \caption{Posterior distributions of $M_{200,\mathrm{crit}}$ for Virgo, Coma, and Perseus across three different inference setups. All three setups produce overlapping posterior mass estimates, though \manticorelocallo yields broader posterior distributions and a slight systematic shift to lower masses for some clusters, particularly Virgo.}
    \label{fig:mass_resolution_convergence}
\end{figure}

\begin{figure}
    \centering
    \includegraphics[width=\columnwidth]{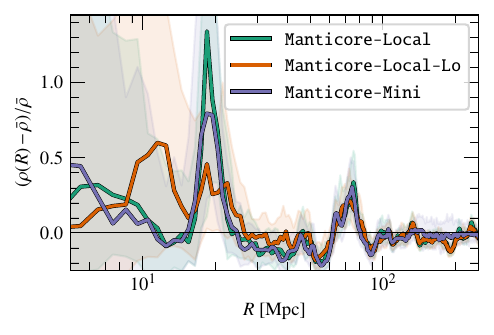}
    \caption{Median radial density profiles of the local supervolume for the three inference setups. All three show excellent agreement in their predicted dark matter distributions out to the edge of their respective constrained regions, with slightly larger fluctuations in the lower resolution run. Shaded regions represent the 10--90 percentile range across posterior resimulations.}
    \label{fig:profile_resolution_convergence}
\end{figure}

To evaluate the robustness of our inference methodology, we compare three inference runs that differ in resolution and the extent of their constrained volumes, while keeping all other components of the model and likelihood fixed. For each we have produced a suite of 50 dark matter-only posterior resimulations at the same particle resolution.

Our fiducial configuration, \manticorelocal, reconstructs a $L = 1000$~Mpc parent volume at $3.9~\mathrm{Mpc}$ resolution, with a constrained region extending to $R = 200~\mathrm{Mpc}$. To test sensitivity to the inference resolution, we repeat the inference at half the inference grid size (\manticorelocallo), using a grid spacing of $7.8~\mathrm{Mpc}$ across the same volume. This allows us to assess convergence in the predicted posterior with respect to the spatial resolution of the inference grid.

A third run, \manticoremini, maintains the high $3.9~\mathrm{Mpc}$ resolution of the fiducial setup but reduces the volume to $L = 500~\mathrm{Mpc}$, with constraints applied only out to $R = 125~\mathrm{Mpc}$. This configuration was designed for more than just convergence testing, but also to explore the impact of constraint extent on structure formation and to enable practical downstream applications.

\Cref{fig:mass_resolution_convergence} presents the posterior distributions of cluster masses for Virgo, Coma, and Perseus. All three inference setups produce consistent predictions, with only minor deviations. As expected, \manticorelocallo yields broader posteriors due to reduced resolving power and displays a slight tendency toward lower peak masses, particularly for Virgo. This is likely due to the difficulty in resolving the survey selection mask and Virgo’s small-scale redshift-space distortions at lower resolution. In contrast, \manticoremini shows excellent agreement with the fiducial \manticorelocal predictions, demonstrating that the reduced constraint radius has little impact on the inferred masses of these prominent clusters.

\Cref{fig:profile_resolution_convergence} compares the radial density profiles inferred by each setup. The three runs yield similar mean profiles, with \manticorelocallo showing slightly larger variance. The agreement across all configurations confirms that the density field reconstruction is stable with respect to both inference resolution and box size.

While \manticorelocallo serves primarily to test resolution convergence, the \manticoremini configuration offers several practical advantages. Its smaller parent box enables higher resolution resimulations and hydrodynamical follow-up runs at lower computational cost. It also facilitates zoom-in simulations of individual structures and provides a framework to study the influence of the constraint radius on the formation history of specific haloes. The strong agreement between \manticoremini and the fiducial setup supports its use as a flexible and efficient tool for future targeted studies of the nearby Universe.

\bsp	
\label{lastpage}
\end{document}